\documentclass[aip,jcp,amsmath,amssymb,reprint]{revtex4-1}

\usepackage{graphicx}
\usepackage{dcolumn}

\usepackage{bm} 
\usepackage{mathptmx} 

\usepackage{xcolor}
\usepackage[version=4]{mhchem}
\usepackage{booktabs}
\usepackage{array}

\newcommand{\ket}[1]{| #1 \rangle}

\newcommand{\braket}[2]{\langle #1 | #2 \rangle}
\newcommand{\matel}[3]{\langle #1 | #2 | #3 \rangle}

\newcommand{\adj}[1]{\ensuremath{\text{adj}} \left( #1 \right)}
\newcommand{\determ}[1]{ \left| #1 \right|}
\newcommand{\trace}[1]{\bigl \langle #1 \bigr \rangle}

\begin{document}

\title{Numerically Stable Resonating Hartree-Fock}

\author{Ericka Roy Miller}
\affiliation{Department of Chemistry, Case Western Reserve University \\ 10900 Euclid Ave, Cleveland, OH 44106, USA}
\author{Shane M. Parker}
\email{shane.parker@case.edu}
\affiliation{Department of Chemistry, Case Western Reserve University \\ 10900 Euclid Ave, Cleveland, OH 44106, USA}
\date{\today}

\begin{abstract}
  The simulation of excited states at low computational cost remains an open challenge
  for electronic structure (ES) methods.
  While much attention has been given to orthogonal ES methods,
  relatively little work has been done to develop
  nonorthogonal ES methods for excited states,
  particularly those involving nonorthogonal orbital optimization.
  We present here a numerically stable formulation of the
  Resonating Hartree-Fock (ResHF) method that uses the matrix adjugate to
  remove numerical instabilities in ResHF
  arising from nearly orthogonal orbitals, and we demonstrate improvements to
  ResHF wavefunction optimization as a result.
  We then benchmark the performance of ResHF against
  Complete Active Space Self-Consistent Field
  in the avoided crossing of LiF,
  the torsional rotation of ethene, and the singlet-triplet energy
  gaps of a selection of small molecules.
  ResHF is a promising excited state method because it incorporates
  the orbital relaxation of state-specific methods, while retaining the
  correct state crossings of state-averaged approaches.
  Our open-source ResHF implementation, yucca, is available on GitLab.
\end{abstract}

\maketitle

\section{\label{sec:intro}Introduction}

One of the most powerful tools for unraveling ultrafast photodynamic processes
is nonadiabatic molecular dynamics (NAMD).
These simulations produce ``molecular movies'', depicting
the response of nuclei and electrons to incident light at femtosecond-to-picosecond
timescales. NAMD simulations have been applied to the study of many diverse
photodynamic processes, from intersystem crossings,\cite{Richter2012JPCL, Penfold2012JCP}
to charge transfers,\cite{Curchod2017JCPA, Nelson2014ACR} to internal conversions
through conical intersections. \cite{Borne2024NC, Schuurman2018ARPC}
The quality of an NAMD simulation is highly dependent on the underlying
electronic structure method selected to provide electronic
energies, nuclear forces, and nonadiabatic coupling vectors.
\cite{Janos2023JCTC, Papineau2024JPCL, Mukherjee2024JCP}
The ideal electronic structure method for use with NAMD simulations
would (i) be computationally efficient,
(ii) generate correct conical intersections, (iii) treat states of different character
with consistent accuracy, and (iv) reliably converge throughout the dynamics.
Unfortunately, finding such a method is challenging, as excited states remain a
frontier of electronic structure method development.

Time-dependent density functional theory (TDDFT) is widely used due to its
favorable cost-to-accuracy ratio for excited states.
However, there are several well-known deficiencies of TDDFT, including a
fundamental neglect of double excitations\cite{Levine2006MP}
and incorrect dimensionality of conical intersections
between ground and excited states.
\cite{Levine2006MP, Taylor2023JCP, Taylor2024JPCA}
A state-averaged (SA) approach to complete active space self-consistent field
(CASSCF) will properly reproduce conical intersections.
However, SA-CASSCF is known to produce low accuracy charge transfer states,
\cite{Segado2016PCCP, Tran2020JPCA}
and perturbative corrections are often applied to remedy this.
Alternatively, a state-specific (SS) approach to
CASSCF has been shown achieve qualitatively accurate charge
transfer states without the need to resort to
expensive perturbative corrections. \cite{Tran2020JPCA}
This is because SS-CASSCF allows for orbital relaxation tailored to an
individual electronic state, while SA-CASSCF is unable to efficiently
accommodate the demands of different electronic environments.
However, excited state SS-CASSCF solutions can be difficult to converge,
and can disappear along a potential energy surface.
\cite{Marie2023JPCA, Saade2024JCTC}
Further, couplings between different electronic states, such as
transition moments or nonadiabatic couplings, are
difficult to obtain from SS solutions.
The dilemma of CASSCF then is its inability to simultaneously achieve
correct conical intersections and qualitatively accurate charge transfer states
without resorting to large active spaces or perturbative corrections.

We posit here that the Resonating Hartree--Fock (ResHF)
\cite{Bremond1964NP,Fukutome1988PoTP} method
resolves this tension between state-specific accuracy
and state-averaged stability.
The ResHF wavefunction is a linear combination of
nonorthogonal Slater determinants. ResHF solutions are
obtained by simultaneously optimizing the orbitals comprising these Slater determinants,
as well as the state expansion coefficients.
ResHF is one of many electronic structure methods
\cite{Burton2019JCTC, Hiberty2002TCA, Lu2022JPCL}
that fall under the general strategy of
nonorthogonal configuration interaction (NOCI).
The unifying mantra of NOCI methods is
``different orbitals for different states''.
NOCI methods have been shown to capture large amounts of
electron correlation with a compact configurational
expansion of the wavefunction.
\cite{Olsen2015JCP, Burton2019JCTC, Sun2024JCTC}
ResHF in particular is able to further reduce the wavefunction size
\cite{Jimenez-Hoyos2013JCP}
and improve excited state energies\cite{Nite2019JCTC}
due to the incorporation of orbital relaxation.
ResHF can be framed as a nonorthogonal analogue of CASSCF,
\cite{Mahler2021JCP} where the constraint of orthogonality
between orbitals from different Slater determinants has been lifted.
One of our main objectives in this contribution is to demonstrate
that the resulting increase in flexibility allows ResHF to
accommodate electronic states of different
character in an unbiased manner via state-averaging.

However, variationally optimizing ResHF wavefunctions is notoriously
difficult,\cite{Jimenez-Hoyos2013JCP, Sun2024JCTC} raising concerns
about the practicality of the method beyond model systems and small
organic molecules.
Recently, Mahler and Thompson shed light on underlying numerical
instabilities present in ResHF. \cite{Mahler2021JCP}
In short, one of the key quantities in ResHF is the
inverse of the interdeterminant overlap matrix, which becomes
singular and thus numerically unstable when two determinants are (nearly) orthogonal. \cite{Fukutome1988PoTP}
Mahler and Thompson addressed this by using generalized functions
to compute orbital derivatives at these points of discontinuity.
\cite{Mahler2021JCP}
In this contribution, we present an alternative solution:
redefining the ResHF density matrix in terms of the
matrix adjugate.\cite{Hill1985SJoAaDM,Stewart1998LAaiA}
The resulting formulation of orbital gradients between
nonorthogonal Slater determinants is consistent with
contributions by Koch and Dalgaard,\cite{Koch1993CPL}
and more recent work by Chen and Scuseria,\cite{Chen2023JCP}
although the approach and context here are distinct.
Our ResHF implementation is available in
our open-source, C++ software package, yucca.
\cite{yucca}

This paper is organized as follows.
In Section \ref{sec:theory}, we review the ResHF method,
detailing the matrix adjugate based reformulation, ResHF-adj.
In Section \ref{sec:methods}, we discuss details of our ResHF implementation
and benchmarking methodology.
In Section \ref{sec:fd} we demonstrate the numerical stability
of ResHF-adj in the context of wavefunction optimization.
In Section \ref{sec:bench}, we benchmark the performance of ResHF against
SA/SS-CASSCF in the avoided crossing of LiF, the torsional rotation of ethene,
and the singlet-triplet energy gaps of a selection of small to medium
molecules from the QUEST database.
We conclude in Section \ref{sec:end} with some
projections about future development of ResHF for excited states.

\section{\label{sec:theory}
Stabilizing Resonating Hartree-Fock with the Matrix Adjugate}

We start by rewriting Fukutome's original ResHF derivation\cite{Fukutome1988PoTP}
with notation based on Burton. \cite{Burton2021JCP}
We refer the interested reader to these
excellent papers for a more thorough treatment of the theory.
Then we examine nonorthogonal orbitals in the limit of orthogonality
as a primary source of numerical instability
and show how the application of the matrix adjugate avoids this issue.
We will use the following notation throughout:
Atomic orbitals (AOs) are designated with indices $\mu \nu \lambda \sigma$,
general molecular orbitals (MOs) are assigned indices $p q r s$, with occupied
MOs $i j k l$ and virtual MOs $a b$.
Slater determinants are labeled with $A B$,
and ResHF electronic states with $I J$.
Matrices are represented using bold font.
By default, these matrices are represented in the AO basis.
Matrices represented in the MO basis are denoted using
subscripted labels: `o' for occupied and `u' for virtual (unoccupied)
MOs.

\subsection{\label{sec:ResHF} Resonating Hartree-Fock (ResHF) Theory}

The ResHF wavefunction,
\begin{equation} \label{eq:reswfn}
  \ket{\Psi_I} = \sum_{A} c_{AI} \ket{\Phi_A} ~ ,
\end{equation}
is a linear combination of nonorthogonal Slater determinants,
$\ket{\Phi_A}$,
where $c_{AI}$ are the state expansion coefficients.
Each Slater determinant has its own set of MOs that we express
in the AO basis as
\begin{equation}
  \phi^A_p(x) = \sum_\mu \chi_\mu(x) C^A_{\mu p},
\end{equation}
where $\chi_\mu$ are spin AO basis functions,
and $C^A_{\mu p}$ are the MO expansion coefficients for Slater determinant $A$.
The MOs within a Slater determinant are orthonormal, while
the MOs in different Slater determinants are allowed
(not required) to be nonorthogonal.

The ResHF energy of state $I$, $E_I$, is defined as
\begin{equation} \label{eq:E}
  E_I = \matel{\Psi_I}{\hat{H}}{\Psi_I}
  = \sum_{AB} c^*_{AI} \matel{\Phi_A}{\hat{H}}{\Phi_B} c_{BI} ,
\end{equation}
where the Hamiltonian operator is
\begin{equation} \label{eq:ham}
  \hat{H} =
  \sum_{\mu\nu} h_{\mu\nu} \hat{a}_\mu^\dagger \hat{a}_\nu
  + \frac{1}{2} \sum_{\mu\nu\lambda\sigma}
  [ \mu \lambda | \nu \sigma ] ~
  \hat{a}_\mu^\dagger \hat{a}_\nu^\dagger \hat{a}_\sigma \hat{a}_\lambda.
\end{equation}
Here, $h_{\mu\nu}$ is a matrix element of the one-electron operator
(containing kinetic energy and nuclear-electron attraction)
and
\begin{equation}
    [ \mu \lambda | \nu \sigma ]
    = \iint d\mathbf{x}_1 \mathbf{x}_2
    \chi^*_\mu (\mathbf{x}_1) \chi_\lambda  (\mathbf{x}_1) \hat{r}^{-1}_{1 2}
    \chi^*_\nu (\mathbf{x}_2) \chi_\sigma (\mathbf{x}_2)
\end{equation}
is a matrix element of the electron-electron repulsion, called electron repulsion integrals (ERI).
In the above, we use square brackets to emphasize that the AOs contain spin degrees of freedom.
In defining the creation and annihilation operators, $\hat{a}_\mu^\dagger$ and $\hat{a}_\mu$,
we follow the notation of Burton. \cite{Burton2021JCP}

We now define several key intermediates used to compute matrix elements between nonorthogonal
Slater determinants. First, $s_{AB} \equiv \langle \Phi_A | \Phi_B\rangle$ is the
interdeterminant overlap, which is the determinant of the overlap matrix between the occupied MOs of
Slater determinants $A$ and $B$,
\begin{equation}
  s_{AB} = \determ{\mathbf{S}^{AB}_{\text{oo}}},
\end{equation}
where
\begin{equation}
  \mathbf{S}^{AB}_{\text{oo}} = \mathbf{C}^{A,\dagger}_{\text{o}} \mathbf{S} \mathbf{C}^{B}_{\text{o}},
\end{equation}
and $\mathbf{S}$ is the AO overlap matrix.
For convenience, our implementation retains a full complement of virtual MOs for each Slater
determinant, in which case $\mathbf{S}^{AB}_{\text{oo}}$ is the occupied-occupied block of the MO overlap matrix,
\begin{equation} \label{eq:Smo}
  \mathbf{S}^{AB}
  = \mathbf{C}^{A,\dagger} \mathbf{S} \mathbf{C}^{B}
  =
  \begin{bmatrix}
    \mathbf{S}^{AB}_{\text{oo}} & \mathbf{S}^{AB}_{\text{ou}} \\
    \mathbf{S}^{AB}_{\text{uo}} & \mathbf{S}^{AB}_{\text{uu}}
  \end{bmatrix} ,
\end{equation}

The interdeterminant one-body density matrix,
$\gamma_{\mu\nu}^{AB} \equiv \langle \Phi_A | \hat{a}_\mu^\dagger \hat{a}_\nu |\Phi_B\rangle$, is
\begin{equation} \label{eq:density}
  \boldsymbol{\gamma}^{AB}
  = s_{AB}  \mathbf{C}^B_o (\mathbf{S}^{AB}_{\text{oo}})^{-1}
  \mathbf{C}^{A,\dagger}_o
  = s_{AB}  \mathbf{Q}^{AB},
\end{equation}
where we have implicitly defined $\mathbf{Q}^{AB}$.
The interdeterminant two-body density matrix is,
\begin{multline} \label{eq:2RDM}
  \Gamma^{AB}_{\lambda\mu\sigma\nu} =
  \matel{\Phi_A}
  {\hat{a}_\mu^\dagger \hat{a}_\nu^\dagger \hat{a}_\sigma \hat{a}_\lambda}{\Phi_B}
  \\
  = s_{AB} \Big[Q^{AB}_{\lambda\mu}Q^{AB}_{\sigma\nu}
  - Q^{AB}_{\sigma\mu}Q^{AB}_{\lambda\nu}\Big] .
\end{multline}
With these definitions, the coupling matrix element, $H_{AB} \equiv \langle \Phi_A | \hat{H} | \Phi_B\rangle$
can be computed as
\begin{equation} \label{eq:hab}
  H_{AB}
  = \trace{\mathbf{h} \boldsymbol{\gamma}^{AB}}
  + \frac{1}{2} \trace{\mathbf{V} \boldsymbol{\Gamma}^{AB}}
  = \trace{\mathbf{h} \boldsymbol{\gamma}^{AB}}
  + \frac{1}{2} \trace{\mathbf{G}^{AB} \boldsymbol{\gamma}^{AB}} ,
\end{equation}
where $\trace{\cdot}$ is the matrix trace, $V_{\mu\lambda\nu\sigma} = [ \mu \lambda | \nu \sigma ]$,
and
\begin{subequations} \label{eq:gab}
  \begin{align}
  G^{AB}_{\mu\nu} &= \sum_{\lambda\sigma} ~
  [ \mu\nu || \lambda\sigma ] ~
  Q^{AB}_{\sigma\lambda} \\
  [ \mu\nu || \lambda\sigma ]
  &= [ \mu\nu | \lambda\sigma ] - [ \mu\sigma | \lambda\nu ] .
  \end{align}
\end{subequations}
We note here that Fukutome\cite{Fukutome1988PoTP} originally
defined $\mathbf{Q}^{AB}$ as the interdeterminant density matrix,
instead of $\boldsymbol{\gamma}^{AB}$.
Similar definitions of $\mathbf{Q}^{AB}$ have been used
by Broer and Nieuwpoort,\cite{Broer1988TCA}
and more recently by
Mahler and Thompson\cite{Mahler2021JCP}
and Burton,\cite{Burton2021JCP} with variant definitions to
handle the cases when $\mathbf{S}^{AB}_{\text{oo}}$ becomes singular.
We omit these variant definitions here,
addressing this issue via the matrix adjugate in Section \ref{sec:res-adj}.

The distinguishing feature of ResHF
is the variational minimization of
the energy with respect to both
the MO expansion coefficients, \{$\mathbf{C}^A$\},
and the ResHF state expansion coefficients, \{$c_{AI}$\}.
To this end, we use a Lagrangian,
\begin{equation} \label{eq:lagrange}
  \mathcal{L} =
  E^{ResHF} - \sum_{IJ} N_{IJ} [\braket{\Psi_I}{\Psi_J} - \delta_{IJ}]
  ,
\end{equation}
where $E^{ResHF}$ can be either single state or state averaged energy and
$N_{IJ}$ is a Lagrange multiplier that enforces
orthonormality of the ResHF electronic states.
The Lagrangian is made stationary when
$N_{IJ} = \delta_{IJ} E_{I} w_I$, where $w_I$ is the weight of state $I$, and
the state expansion coefficients satisfy
the generalized eigenvalue equation,
\begin{equation} \label{eq:eigen}
  \mathbf{H} \mathbf{c} = \mathbf{s} \mathbf{c} \mathbf{E},
\end{equation}
where $\mathbf{H}$, $\mathbf{c}$, and $\mathbf{s}$ are the matrix representations of
the determinant coupling, state expansion coefficients, and determinant overlaps,
respectively, and $\mathbf{E}$ is a diagonal matrix of the energy eigenvalues.
To optimize the MOs,
we build a set of Resonating Fock matrices, shown here in the AO basis,
\begin{subequations} \label{eq:fock}
  \begin{align}
    \mathbf{F}^A &= \sum_I w_I \mathbf{F}^A_I \\
    \mathbf{F}^A_I &=
    \mathbf{F}^{AA} |c_{AI}|^2
    + \sum_{B \neq A} c_{AI}^*\left[ \mathbf{K}^{AB}
    + (\mathbf{K}^{AB})^{\dagger} \right] c_{BI}
    ,
  \end{align}
\end{subequations}
where we have defined the interdeterminant Fock matrix,
$\mathbf{F}^{AB} = \mathbf{h} + \mathbf{G}^{AB}$.
The $\mathbf{K}^{AB}$ term can be written in the AO basis as
\begin{multline} \label{eq:kabao}
  \mathbf{K}^{AB} =
  s_{AB}
  (\mathbf{1} - \mathbf{S} ~ \mathbf{Q}^{AB}) ~
  \mathbf{F}^{AB} ~ \mathbf{Q}^{AB} ~ \mathbf{S} \\
  + (H_{AB} - s_{AB} E_I) ~
  \mathbf{S} ~ \mathbf{Q}^{AB} ~ \mathbf{S} .
\end{multline}
Recognizing that $\mathbf{Q}^{AB}$ is an oblique projector that maps
occupied orbitals in $A$ to occupied orbitals in $B$,
\begin{equation}
  \mathbf{Q}^{AB} \mathbf{S} \mathbf{Q}^{AB} = \mathbf{Q}^{AB},
\end{equation}
we see that $\mathbf{K}^{AB}$ resembles a virtual-occupied projected Fock matrix.
We conclude this section by observing the above ResHF
equations simplify to the Hartree--Fock equations in the single Slater determinant case.

\subsection{\label{sec:res-adj}
  Overcoming Numerical Instability with the Matrix Adjugate}

A fundamental question that arises when evaluating matrix elements
involving nonorthogonal Slater determinants is what happens when the
overlap between Slater determinants approaches zero?
This question is particularly pressing for ResHF in particular,
because orbital relaxation tends to push overlaps to be small
but nonzero. \cite{Nite2019JCTC}
For a completely orthogonal pair of Slater determinants,
$\mathbf{S}^{AB}_{\text{oo}}$ is singular, meaning its inverse does not exist.
Even when the overlap is small but nonzero, the condition number of
$\mathbf{S}^{AB}_{\text{oo}}$ becomes large, leading to numerical instability.
Moreover, in the (near) singular case, the orbital gradients can become ill-defined,
which would seem to preclude reliable convergence.
From our perspective, this issue is an artifact of factoring out common
intermediates like the determinant overlap.

The origin of numerical instability in ResHF can be illustrated with
the singular value decomposition (SVD) of $\mathbf{S}^{AB}_{\text{oo}}$,
\begin{equation} \label{eq:svd}
  \mathbf{S}^{AB}_{\text{oo}}
  = \mathbf{U}^{AB} ~ \boldsymbol{\Sigma}^{AB}
  ~ \mathbf{V}^{AB,\dagger},
\end{equation}
where $\mathbf{U}^{AB}$ and $\mathbf{V}^{AB}$ are unitary matrices,
and $\boldsymbol{\Sigma}^{AB}$ is a diagonal matrix with singular values $\sigma_i$.
As two Slater determinants become orthogonal, at least one singular value, $\sigma_i$,
approaches zero, and the inverse, which contains $1/\sigma_i$, diverges.
A common solution to this problem is to use different expressions according
to the number of (near) zero singular values. \cite{Mahler2021JCP}
However, this approach leads to discontinuities in potential energy surfaces
as the number of (near) zero singular values changes, and thus we do not further
consider it here.
An alternative approach is to apply a floor to the singular values using a
user-selected cutoff, $\epsilon$, such that
$\sigma_i = \max(\sigma_i, \epsilon)$.
We will refer to this implementation strategy as ResHF-cutoff.
Compared to the previous solution, the cutoff approach allows for some
control of the numerical stability, should produce smooth potential energy surfaces,
and is relatively straightforward to implement, albeit with a difficult-to-control
loss of accuracy.

We argue that a better route is to reformulate ResHF in terms of the matrix adjugate,
which can be defined for nonsingular matrices as
\cite{Hill1985SJoAaDM,Stewart1998LAaiA}
\begin{equation}
  \adj{\mathbf{M}} \equiv \determ{\mathbf{M}} (\mathbf{M})^{-1}.
\end{equation}
Remarkably, even in the limit of singular $\mathbf{M}$ the matrix adjugate remains well-defined.
To clarify the connection to ResHF, recall that $s_{AB} = \determ{\mathbf{S}^{AB}_{\text{oo}}}$,
which lets us rewrite the ResHF density matrix in terms of the matrix adjugate,
\begin{equation}
  \boldsymbol{\gamma}^{AB}
  = \mathbf{C}^B_{\text{o}} ~ \adj{\mathbf{S}^{AB}_{\text{oo}}} ~
  \mathbf{C}^{A,\dagger}_{\text{o}}.
\end{equation}
One of the key results of this paper is that ResHF energies and gradients can be
rewritten in a numerically stable form using the matrix adjugate.

The matrix adjugate can be efficiently computed using the SVD
of $\mathbf{S}^{AB}_{\text{oo}}$,
\begin{equation}
  \adj{\mathbf{S}^{AB}_{\text{oo}}} = \eta \mathbf{V}^{AB} ~ \boldsymbol{\Xi} ~ \mathbf{U}^{AB,\dagger},
\end{equation}
where $\boldsymbol{\Xi}$ is a diagonal matrix with elements
\begin{equation} \label{eq:xi}
  \xi_i = \prod_{j \neq i} \sigma_j ,
\end{equation}
and $\eta = \determ{\mathbf{U}^{AB} (\mathbf{V}^{AB})^{\dagger}} = \pm 1$.
Here, we see why the matrix adjugate is well-defined in the limit of singular $\mathbf{S}^{AB}_{\text{oo}}$:
the product of singular values contained in the matrix determinant cancels the division by singular value
analytically, thus avoiding a numerical division by zero.

Using the above, $\boldsymbol{\gamma}^{AB}$ can be rewritten as
\begin{equation} \label{eq:gamma-adj}
  \boldsymbol{\gamma}^{AB} = \eta ~
  \mathbf{B}_{\text{o}} \boldsymbol{\Xi}^{AB} \mathbf{A}_{\text{o}}^{\dagger} ,
\end{equation}
where
\begin{subequations}
  \begin{align}
    \mathbf{A}_{\text{o}} &= \mathbf{C}^A_{\text{o}} ~ \mathbf{U}^{AB} \\
    \mathbf{B}_{\text{o}} &= \mathbf{C}^B_{\text{o}} ~ \mathbf{V}^{AB}
  \end{align}
\end{subequations}
form a biorthogonal basis.
We note that the matrices $\mathbf{A}_o$ and $\mathbf{B}_o$ are of use only for the specific
$AB$ pair, but we omit $AB$ superscripts for brevity.

To stabilize the ResHF energy, Eq. \eqref{eq:E},
using ResHF-adj, we turn our attention to the determinant coupling, $H_{AB}$.
The one-electron contribution to $H_{AB}$ can be straightforwardly stabilized by
using Eq. \eqref{eq:gamma-adj} for the density matrix.
For the two-electron contribution to $H_{AB}$,
we express the ERIs in the SVD MO basis representation to get
\begin{equation}
  \frac{1}{2} \trace{\mathbf{G}^{AB} ~ \boldsymbol{\gamma}^{AB}}
  = \eta \frac{1}{2}
  \sum_{ij}
    [i_A i_B || j_A j_B]
  \xi_{ij} ,
\end{equation}
where
\begin{equation}
  \xi_{ij} = \frac{1}{\sigma_i} \frac{1}{\sigma_j} \prod_{m} \sigma_m.
\end{equation}
Because of the diagonal terms, $\xi_{ii}$, this expression appears to be
numerically unstable.
However, the diagonal case here corresponds to self-interaction, which is exactly canceled.
That is, $[i_Ai_B||i_Ai_B] = 0$, and thus $\xi_{ii}$ will not contribute to the energy.
We therefore define $\xi_{ij}$ as
\begin{equation} \label{eq:Xij}
  \xi_{ij}
  =
  \begin{cases}
    \prod_{m \ne i,j} \sigma_m  & \text{if } i \ne j \\
    \tau & \text{if } i = j
  \end{cases} ,
\end{equation}
where $\tau$ is a parameter. The cancellation of all terms involving $\xi_{ii}$
means that the result should not depend on what value is used for $\tau$.
Indeed, we use this as a test of our implementation, and verify that the energies
and gradients are identical for different values of $\tau$.
We default to setting $\tau = 1$.

We now turn our efforts to
the Resonating Fock matrix, focusing on $\mathbf{K}^{AB}$.
In the ResHF-cutoff approach, the $\mathbf{K}^{AB}$ matrix
can be computed directly in the AO basis (Eq. \eqref{eq:kabao}),
significantly simplifying the implementation.
For ResHF-adj, we compute
$\mathbf{K}^{AB}$ in the SVD transformed MO basis,
\begin{multline} \label{eq:kabmo}
  \mathbf{A}^\dagger \mathbf{K}^{AB} \mathbf{A} = \\
  \begin{bmatrix}
    \mathbf{A}_{\text{o}}^\dagger \mathbf{K}^{AB} \mathbf{A}_{\text{o}} &
    \mathbf{A}_{\text{o}}^\dagger \mathbf{K}^{AB} \mathbf{A}_{\text{u}}  \\
    \mathbf{A}_{\text{u}}^\dagger \mathbf{K}^{AB} \mathbf{A}_{\text{o}} &
    \mathbf{A}_{\text{u}}^\dagger \mathbf{K}^{AB} \mathbf{A}_{\text{u}}
    \end{bmatrix}
    =
    \begin{bmatrix}
      \mathbf{K}^{AB}_{\text{oo}} &
      \mathbf{K}^{AB}_{\text{ou}}  \\
      \mathbf{K}^{AB}_{\text{uo}} &
      \mathbf{K}^{AB}_{\text{uu}}
    \end{bmatrix} ,
\end{multline}
where $\mathbf{A}_{\text{u}} = \mathbf{C}^A_{\text{u}}$ and $\mathbf{A}_{\text{o}} = \mathbf{C}^A_{\text{o}} \mathbf{U}^{AB}$.
The $\mathbf{K}^{AB}_{\text{ou}}$ and $\mathbf{K}^{AB}_{\text{uu}}$ block matrices
vanish because
\begin{equation}
  \mathbf{Q}^{AB} \mathbf{S} \mathbf{A}_{\text{u}}
  = \mathbf{B}_{\text{o}} (\boldsymbol{\Sigma}^{AB})^{-1}
  \mathbf{A}_{\text{o}}^\dagger \mathbf{S} \mathbf{A}_{\text{u}} = 0 .
\end{equation}
For the $\mathbf{K}^{AB}_{\text{oo}}$ block matrix, we find
\begin{subequations}
  \begin{align}
    \mathbf{A}_{\text{o}}^\dagger (\mathbf{1} - \mathbf{S} \mathbf{Q}^{AB})
    &= \mathbf{A}_{\text{o}}^\dagger - \boldsymbol{\Sigma}^{AB}
    (\boldsymbol{\Sigma}^{AB})^{-1}
    \mathbf{A}_{\text{o}}^\dagger = 0
    \\
    \mathbf{A}_{\text{o}}^\dagger \mathbf{S} \mathbf{Q}^{AB} \mathbf{S} \mathbf{A}_{\text{o}}
    &= \boldsymbol{\Sigma}^{AB} (\boldsymbol{\Sigma}^{AB})^{-1}
    \mathbf{A}_{\text{o}}^\dagger \mathbf{S} \mathbf{A}_{\text{o}}
    = \mathbf{1} ,
  \end{align}
\end{subequations}
where $\mathbf{1}$ is an $n_e\times n_e$ identity matrix, where $n_e$ is the number of electrons.
Therefore, the only surviving terms are
\begin{equation}
  \mathbf{K}^{AB}_{\text{oo}} = (H_{AB} - s_{AB} E_I)
  \mathbf{1}.
\end{equation}

Starting from
\begin{multline} \label{eq:kvo}
  \mathbf{K}^{AB}_{\text{uo}} =
  s_{AB}
  \mathbf{A}_{\text{u}}^\dagger ~
  (\mathbf{1} - \mathbf{S} ~ \mathbf{Q}^{AB})
  \mathbf{F}^{AB} ~ \mathbf{Q}^{AB} ~ \mathbf{S}
  ~ \mathbf{A}_{\text{o}} \\
  + (H_{AB} - s_{AB} E_I)
  \mathbf{A}_{\text{u}}^\dagger ~
  \mathbf{S} ~ \mathbf{Q}^{AB} ~ \mathbf{S}
  ~ \mathbf{A}_{\text{o}} ,
\end{multline}
we group terms according to the number of inverse
MO overlap matrices present to get
\begin{align} \label{eq:Kai}
  K^{AB}_{ai} &=
  \eta \Big(
    \bar{h}^{AB}_{ai} - E_I \bar{S}^{AB}_{ai}
    \Big)
  \xi_i \nonumber
  \\
  &+ \eta \sum_j \Big(
    \bar{S}^{AB}_{ai} \bar{h}^{AB}_{jj}
    - \bar{S}^{AB}_{aj} \bar{h}^{AB}_{ji}
    + [a_A i_B | | j_A j_B]
    \Big)
  \xi_{ij}
  \\ \nonumber
  &+ \eta \sum_{jk} \Big(
      \frac{1}{2} \bar{S}^{AB}_{ai} [j_A j_B | | k_A k_B]
      - \bar{S}^{AB}_{aj} [j_A i_B | | k_A k_B]
    \Big)
  \xi_{ijk} ,
\end{align}
where the one-electron potential and AO overlap matrices have
been transformed into the SVD MO representation as
\begin{align}
  \bar{h}^{AB}_{pq}
  &= \sum_{\mu\nu} A^*_{\mu p} h_{\mu\nu} B_{\nu q}
  \\
  \bar{S}^{AB}_{pq}
  &= \sum_{\mu\nu} A^*_{\mu p} S_{\mu\nu} B_{\nu q} ,
\end{align}
and we introduced $\xi_{ijk}$ as
\begin{equation}
  \xi_{ijk} = \frac{1}{\sigma_i} \frac{1}{\sigma_j} \frac{1}{\sigma_k}
  \prod_{m} \sigma_m
 .
\end{equation}

The first term in $K^{AB}_{ai}$ is stable because $\xi_i$ has no singular values in the denominator.
The second term is stable because the summand vanishes when $i = j$, thus allowing us to use
Eq. \eqref{eq:Xij}.
The third term is also stable, but more challenging to see, so we will
show it directly. To show that the third term is stable, it suffices to show that
the summand will vanish when any two indices of $\xi_{ijk}$ are equal, because we
can then define $\xi_{ijk}$ in analogy to Eq. \eqref{eq:Xij}.
First, we see that the summand vanishes when $j = k$, because the antisymmetrized
integrals vanish.
Next, we symmetrize the summand with respect to $j$ and $k$, and use this
to restrict the summation to be
\begin{multline} \label{eq:Kijk}
\sum_{j<k}
  \Big(
      \bar{S}^{AB}_{ai} [j_A j_B | | k_A k_B]
      \\
      - \bar{S}^{AB}_{aj} [j_A i_B | | k_A k_B]
      - \bar{S}^{AB}_{ak} [k_A i_B | | j_A j_B]
  \Big)
  \xi_{ijk},
\end{multline}
where we have used the symmetry of the ERIs and $\xi_{ijk}$ to simplify.
Now we can see that when $j=i$, the first two terms of Eq. \eqref{eq:Kijk} cancel,
and the third vanishes. Similarly, when $k=i$, the first and third terms cancel,
and the second vanishes. Thus, every term in $K^{AB}_{ai}$ is numerically stable.
As a result, we redefine $\xi_{ijk}$ as
\begin{gather} \label{eq:Xijk}
  \xi_{ijk} =
  \begin{cases}
    \prod_{m \ne i,j,k} \sigma_m  & \text{if } i \ne j \ne k \\
    \tau & \text{otherwise}
  \end{cases} ,
\end{gather}
where we use the same $\tau$ is in Eq. \eqref{eq:Xij}.

A major consequence of using this formulation
is that it requires the ERIs to be computed in an MO basis,
whereas Hartree--Fock implementations are often most efficient (partially) in the
AO basis.
To avoid a costly 4-index transformation of the ERIs, we use the resolution of the identity (RI)
approximation\cite{Baerends1973CP,Dunlap79JChemPhys} for all ERIs,
\begin{equation} \label{eq:df}
  (p q | r s)
  \approx \sum_{PQ}
  (p q | P ) (P | Q)^{-1} (Q | r s) ,
\end{equation}
where $(pq|P)$ and $(P|Q)$ are 3-center and 2-center ERIs, respectively,
and  $PQ$ label auxiliary fitting basis functions.
With the RI approximation, the overall scaling of the ResHF-adj method is
$\mathcal{O}(N^5)$, where $N$ is a measure of system size.
By contrast, Hartree--Fock with the RI approximation scales as $\mathcal{O}(N^4)$.
\cite{Weigend2002PCCP}
The increase in scaling comes from the $\xi_{ijk}$ terms, which
complicate separation of terms.
A detailed discussion of how RI was used to implement Eq. \eqref{eq:Kai}
is provided in Appendix \ref{sec:df}.

\section{\label{sec:methods}
Methodology and Implementation Details}

The matrix adjugate formulation of ResHF was implemented into our open-source
code, yucca.
\cite{yucca}
All SA-CASSCF calculations were performed using ORCA version
5.0.0. \cite{Neese2022WIREs}
SS-CASSCF calculations for LiF were also performed with ORCA,
while SS-CASSCF energies for ethene were obtained
from Reference \onlinecite{Saade2024JCTC}.
All input files and data can be found in an Open Science Framework
repository. \cite{OSFadj}

Default convergence criteria
($10^{-7} ~ E_h$ energy change and
$10^{-3} ~ E_h$ orbital gradient)
were used for all CASSCF calculations performed in ORCA.
The first-order perturbative SuperCI method\cite{Kollmar2019JCC} was used
to converge the CASSCF orbitals.
To match the CASSCF convergence criteria,
ResHF calculations were considered converged when the state-averaged
energy changed by less than
$10^{-7} ~ E_h$ and the orbital gradient was below
$10^{-3} ~ E_h$, unless stated otherwise.
The direct inversion of the iterative subspace (DIIS)\cite{Pulay1982JCC}
in combination with the maximum overlap method was
used to converge the ResHF orbitals.
For both the ethene and LiF scans, converged ResHF and CASSCF orbitals
from the prior step were used as initial guesses for their respective methods.

The def2-SVP basis set\cite{Weigend2005PCCP}
was used in both the numerical stability
analysis and LiF bond dissociation.
The aug-cc-pVDZ basis set\cite{Dunning1989JCP, Kendall1992JCP}
was used to compute the ethene torsional curve in order to be consistent
with SS-CASSCF energies from Reference \onlinecite{Saade2024JCTC}.
Vertical excitation energies of QUEST database molecules were computed
using the def2-QZVP basis set. \cite{Weigend2003JCP}

All ResHF calculations were performed using unrestricted determinants.
Detailed ResHF equations using unrestricted MOs can be found
in Appendix \ref{sec:ureshf}.

\begin{figure}[htbp]
  \includegraphics[width=\linewidth]{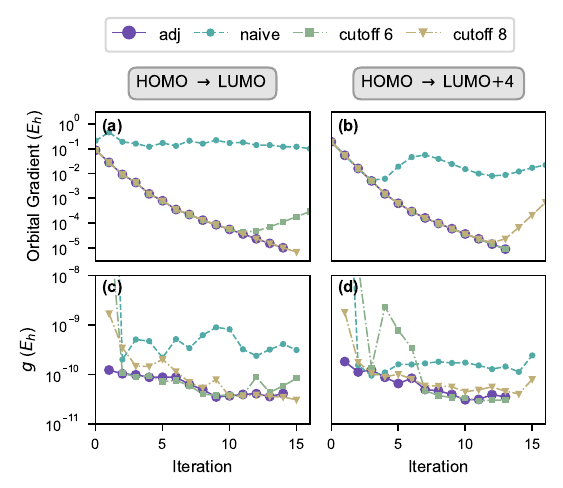}
  \caption{\label{fig:fderr}
  Orbital gradient (a, b) and minimum Fock matrix error, $g$ (c, d)
  as a function of iteration number for planar ethene using 3SA-ResHF(3sd)/def2-SVP.
  Two ResHF wavefunctions are compared: initialized from a HOMO
  $\rightarrow$ LUMO single excitation (a, c) and
  initialized from a HOMO $\rightarrow$ LUMO+4 single excitation (b, d).
  A tight orbital gradient threshold
  ($10^{-5} ~ E_h$) was used.
  Consistent convergence and low Fock matrix errors are seen with
  matrix adjugate stabilized ResHF (purple), while
  unstabilized ``naive'' ResHF (teal) fails to converge due to large
  errors in the Fock matrix. ResHF with singular value cutoffs
  of $10^{-6}$ (green) and $10^{-8}$ (tan)
  perform inconsistently between ResHF wavefunction initializations.
  }
\end{figure}

\section{\label{sec:fd}
Numerical Analysis of ResHF Algorithm Stability}

To begin the analysis, we first compare the wavefunction optimization
performances of the ResHF-adj and ResHF-cutoff implementations, using
planar ethene.
We compare two ResHF wavefunctions, each built from 3 Slater determinants:
the HF ground state, an alpha-spin single excitation,
and a beta-spin single excitation.
For the first ResHF wavefunction, we initialize the determinants from single excitations from
HOMO $\rightarrow$ LUMO, while the second ResHF wavefunction is built
from HOMO $\rightarrow$ LUMO$+4$ single excitations, where HOMO and LUMO stand for
highest occupied MO and lowest unoccupied MO, respectively.
This initial guess ensures that Slater determinants are orthogonal
for the initial wavefunction,
meaning numerical instabilities are expected from the onset.

Three ResHF strategies are used to attempt to converge these wavefunctions:
(i) a naive approach where no attempts at
mitigating numerical instability are made,
(ii) the ResHF-cutoff approach, where a user set threshold, $\epsilon$,
is applied to small singular values,
and (iii) the ResHF-adj approach.
The progress of the optimizations are shown in
Figure \ref{fig:fderr}a and \ref{fig:fderr}b.
The naive approach dramatically fails to converge
for either the HOMO $\rightarrow$ LUMO or
the HOMO $\rightarrow$ LUMO$+4$ initial guesses.
ResHF-adj, on the other hand, converges monotonically for both initial guesses.
With the right cutoff, similar convergence behavior can
be achieved using the ResHF-cutoff approach. However, we find that different
initial guesses require different cutoffs for stable convergence. For example,
in the HOMO $\rightarrow$ LUMO
case, $\epsilon=10^{-8}$ converges,
while $\epsilon=10^{-6}$ fails to converge
(Figure \ref{fig:fderr}a). Yet we observe the opposite in the
HOMO $\rightarrow$ LUMO$+4$ case: $\epsilon=10^{-6}$
converges, while $\epsilon=10^{-8}$ does not (Figure \ref{fig:fderr}b).

The differences in convergence performance can be understood from
examining the errors present in the Resonating Fock matrices
throughout the optimization process, as shown in
Figures \ref{fig:fderr}c and \ref{fig:fderr}d.
We assess the accuracy of the off-diagonal blocks of the Resonating Fock matrices
by comparing to numerical orbital gradients obtained through finite difference,
quantifying the error as
\begin{equation}
  g_{err} (h) =
  \sqrt{
    \sum_{Aia} |
    F^{A}_{ia} - \frac{d E^{SA}}{d \kappa^{A}_{ai}}(h)
    |^2
  },
\end{equation}
where $\kappa^{A}_{ai}$ is an orbital rotation between an
occupied ($\phi^A_i$) and virtual ($\phi^A_a$) spin MO pair,
and $h$ represents the finite difference step size.
Numerical orbital gradients were computed with a fourth order
central finite difference stencil.
To isolate the error in the Fock matrix from the error of the finite
difference approximation, we compute the numerical gradient using step sizes ranging
from $10^{-4}$ to $10^{-1}$, and take the minimum, $ g = \min_h g(h)$, as a proxy for
the quality of the analytical gradients.
We note that we use the matrix adjugate stabilized energy expression
to compute the finite difference energies in all cases.

The approximate errors, $g$, are plotted
for each SCF iteration in Figure \ref{fig:fderr}c and \ref{fig:fderr}d.
ResHF-adj shows a \emph{consistently} low level of error,
sitting at or below $1.8 \times 10^{-10} ~ E_h$ for each step
of the optimization. The ResHF-cutoff approaches start off with
several orders of magnitude larger errors than ResHF-adj at the first iteration, but
significantly reduce these errors as the optimization
progresses. However, there are seemingly random spikes in error
for ResHF-cutoff throughout optimization. While these error spikes might
appear minimal, they nevertheless can preclude convergence
(Figure \ref{fig:fderr}a and \ref{fig:fderr}b).

\begin{figure}[htbp]
  \includegraphics[width=\linewidth]{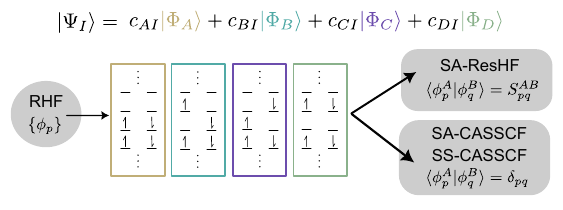}
  \caption{\label{fig:wfn}
  Wavefunction initialization for ResHF and CASSCF methods.
  Initial orbitals were obtained from a restricted Hartree-Fock calculation,
  then used to build a 4 determinant active space.
  Three modes of orbital relaxation were then applied:
  SA-ResHF, SA-CASSCF, and SS-CASSCF.
  In CASSCF orbitals between determinants are constrained to
  remain orthonormal during orbital relaxation,
  while this constraint is lifted for ResHF.}
\end{figure}

\section{\label{sec:bench}
Benchmarking Excited State Resonating Hartree-Fock Performance}

In this section, we seek to examine the performance of ResHF against CASSCF.
To orient the reader, we summarize here some key differences between
these methods.
In SS-CASSCF, a single set of molecular orbitals
is optimized for a single electronic state. Because one
set of orthonormal orbitals are used, the Slater determinants within
the complete active space are orthonormal. However, SS-CASSCF
solutions for different electronic states are not orthogonal
because they are the result of separate optimizations.
Orthogonal electronic states can be recovered via state interaction
\cite{Malmqvist1989CPL},
but this requires the evaluation of matrix elements between
nonorthogonal Slater determinants.
By contrast, SA-CASSCF ensures orthogonal electronic
states by optimizing orbitals for an average of the electronic states.
This can come at the cost of accuracy when using SA-CASSCF to describe
electronic states of different character, such as charge transfer states.
\cite{Segado2016PCCP, Tran2020JPCA}
In our view, SA-ResHF is intermediate between SS-CASSCF and SA-CASSCF in the sense
all electronic states are optimized simultaneously, but
each Slater determinant has its own set of orbitals.
Put another way, SA-ResHF can be obtained by removing the
constraint of using a single set of orbitals from SA-CASSCF.

In the following,
we show that SA-ResHF is about as accurate as SS-CASSCF,
without sacrificing the benefits of state-averaging.
We begin by examining the performance of SA-ResHF and SA/SS-CASSCF for charge
transfer states with the bond dissociation of LiF,
before examining the continuity of excited state surfaces
along the torsional rotation of ethene.
We will then apply minimal sized SA-ResHF and SA-CASSCF wavefunctions
to compute singlet-triplet energy gaps of a selection of small
molecules.

\begin{figure*}[htbp]
  \includegraphics[width=\linewidth]{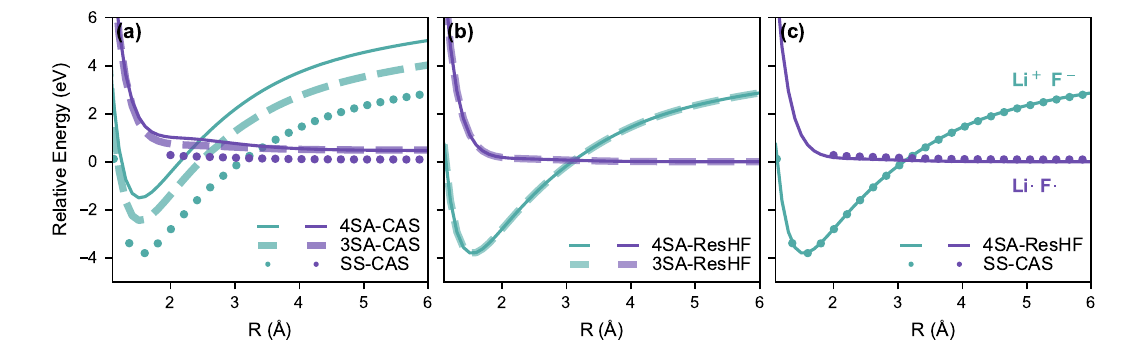}
  \caption{\label{fig:lif}
  The first two singlet states of the LiF bond dissociation.
  Energies are displayed relative to the UHF energy of LiF at
  a fully dissociated bond length (8.0 $\AA$).
  All energies were computed using the def2-SVP basis set.
  Panel (a) compares SS-CASSCF(2e,2o) to 3SA and 4SA-CASSCF(2e,2o),
  panel (b) compares 3SA and 4SA-ResHF(4sd) energies,
  and panel (c) compares SS-CASSCF(2e,2o) and 4SA-ResHF(4sd).
  }
\end{figure*}

To allow for direct comparison between the methods, the same sized
wavefunction was used for ResHF and CASSCF, unless otherwise specified. A sketch of the
wavefunction initialization scheme is presented in Figure \ref{fig:wfn}.
Both CASSCF and ResHF initial guesses for molecular orbitals were obtained from
restricted Hartree-Fock (RHF) calculations, unless otherwise specified.
For SS/SA-CASSCF, a (\textit{2e, 2o}) active space was used,
while a set of 4 Slater determinants were used to build the ResHF wavefunction.
To mimic the full configuration interaction expansion of the CASSCF active space,
the ResHF determinants were initialized as the closed-shell RHF determinant,
a single alpha-spin electron excitation out of the RHF determinant,
the paired beta-spin single electron excitation, and a double electron excitation
(Figure \ref{fig:wfn}).
We will use the ResHF-adj implementation exclusively in the
following analysis.

\subsection{\label{sec:lif}
Flexible ResHF Orbital Relaxation in LiF Dissociation}

The LiF bond dissociation is a useful model to evaluate how well
electronic structure methods
can accommodate the contradictory demands of charge transfer
and locally excited states.
Marie and Burton\cite{Marie2023JPCA} have show that several SS-CASSCF
solutions exist for this system.
We target two electronic states, one ionic and
one neutral. This pair of states was referred to as ``quasi-diabatic''
by Marie and Burton\cite{Marie2023JPCA} because they exhibit a weakly avoided crossing.
The \ce{S0} state has ionic character at bond equilibrium
and mixes with the diradical \ce{S1} state as the LiF bond stretches.
\cite{Kahn1974JCP, Bauschlicher1988JCP, Marie2023JPCA}
A state-averaged approach in CASSCF will attempt to describe both states
with a single set of orbitals, which is challenging for a small active space.
We also include
the lowest lying triplet state in the state-averaging.
Therefore, SA-CASSCF will tend to favor the diradical state over the
ionic state during orbital relaxation, while SS-CASSCF will be free from this bias.

Applying the 4 determinant ResHF/CASSCF wavefunctions (Figure \ref{fig:wfn}),
we compute the LiF bond
dissociation curve for the ionic and covalent singlet states.
SA-CASSCF and SA-ResHF curves were generated by starting at a bond length of
8.0 \AA{} and scanning at an interval of 0.1 \AA.
We ensured determinants were initialized to include the p orbital on the F
atom that aligned with the LiF bond.
For state-averaging schemes, we applied a 3SA equal weighting
across ionic, covalent, and \ce{T1} states and a 4SA equal weighting
that also includes the high lying \ce{S2} state.
The SS-CASSCF covalent state was also obtained by
starting at a fully dissociated 8.0 $\AA$ structure and using RHF MOs
in the initial guess.
Convergence failures occurred for the SS-CASSCF covalent state scan
at lengths shorter than 2.0 \AA.
We attribute this to limitations of the default convergence routines we selected
in ORCA, rather than a disappearance of the quasi-diabatic SS-CASSCF solution. \cite{Marie2023JPCA}
To obtain the SS-CASSCF ionic state, the scan was started at a bond length
of 1.6 \AA. We also found it necessary to initialize the SS-CASSCF ionic
wavefunction using the 3SA-CASSCF converged orbitals in order to ensure
the ionic state did not collapse to the covalent state near the crossing.
For clarity, we have omitted the lowest lying \ce{T1} state
and the high energy \ce{S2} state from Figure \ref{fig:lif}.

We compare the SA-CASSCF and SS-CASSCF potential energy surfaces in
Figure \ref{fig:lif}a.
SS-CASSCF, where orbitals are optimized to exclusively accommodate
either the ionic or covalent state, yields the lowest energy surfaces
and indicates an crossing near 3.1 \AA.
The 3SA-CASSCF approach raises the ionic state bond equilibrium
energy by 1.4 eV and shifts the crossing point to 2.7 \AA.
Introducing the high energy \ce{S2} state into the average (4SA-CASSCF)
further deteriorates the quality of the potential energy surfaces,
raising the bond equilibrium energy by 2.4 eV
and shifting the crossing point to 2.5 \AA.
In contrast, ResHF is insensitive to the number of states included
in state-averaging (Figure \ref{fig:lif}b)
and almost exactly reproduces the SS-CASSCF potential energy surfaces (Figure \ref{fig:lif}c).
Thus, ResHF allows for tailored orbital relaxation for individual electronic states,
while preserving the advantages of state averaging.

\subsection{\label{sec:ethene}
Well-Defined ResHF Surfaces in Ethene Torsional Scan}

As powerful as an SS-CASSCF approach can be for excited states,
a well-known issue is the disappearance of individual solutions along a
potential energy surface.
Recently, Saade and Burton\cite{Saade2024JCTC} have performed extensive
analysis of the excited SS-CASSCF(2e, 2o) solutions of ethene during torsional rotation
about the carbon--carbon double bond.
Here, we compare the performance of SS-CASSCF in ethene to SA-ResHF and SA-CASSCF,
to demonstrate the stability of a state-averaged approach.

With the CASSCF(2e, 2o) and ResHF(4sd) wavefunctions, we are able to model
three low-lying excited states of interest: the triplet
$\pi \rightarrow \pi^*$ excitation, which is commonly denoted as T in
Mulliken labeling, the singlet $\pi \rightarrow \pi^*$ single excitation (V),
and the double $(\pi)^2 \rightarrow (\pi^*)^2$ excitation ionic state (Z).
The ionic Z state, with a vertical excitation energy around 13 eV at a planar geometry,
dramatically reduces in energy as ethene undergoes torsional rotation,
becoming strongly coupled to the
V state at the 90-degree torsional rotation. \cite{Ben-Nun2000CP}
The V state is a particularly challenging state to accurately model
owing to its much more diffuse nature than its triplet counterpart, the T state.
\cite{Feller2014JCP}

\begin{figure}[htbp]
  \includegraphics[width=\linewidth]{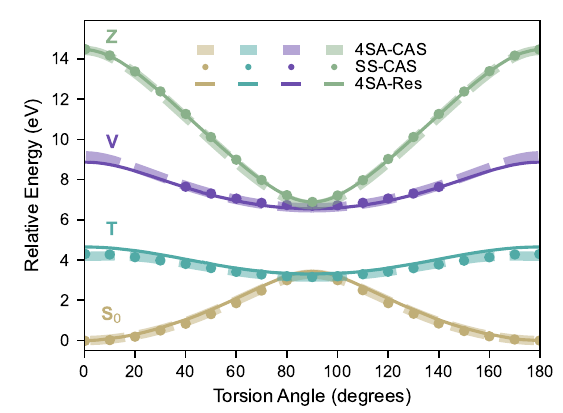}
  \caption{\label{fig:ethene}
  Torsional scan about the carbon-carbon bond of ethene, showing the singlet
  ground state and the T, V, and Z excited states.
  4SA-ResHF(4sd) and 4SA-CASSCF(2e,2o) energies were computed using the aug-cc-pVDZ basis set.
  SS-CASSCF(2e,2o) energies were digitized from Reference \onlinecite{Saade2024JCTC}.
  Energies are displayed as relative to the \ce{S0} energy of planar ethene computed
  by the respective electronic structure method.
  }
\end{figure}

Starting from the 90 degree twisted structure,
we initialize the ResHF determinants with single and double excitations
from the highest occupied and lowest unoccupied RHF molecular orbitals.
An equal weight state averaging scheme was applied
across the three singlet states and one triplet state for both
CASSCF and ResHF.
From the 90 degree structure, we scan towards planar ethene at an interval of
0.05 degrees.
The resulting energy surfaces are shown in Figure \ref{fig:ethene}.

Little deviation is observed between SA-ResHF, SA-CASSCF, and SS-CASSCF
potential energy surfaces. The largest deviation between SA-CASSCF
and SA-ResHF is observed at the planar ethene structure,
with SA-ResHF coming in 0.5 eV higher in energy for the T state and
0.3 eV lower for the V state, relative to SA-CASSCF.
SA-ResHF and SA-CASSCF produce qualitatively similar, yet visually distinct
electronic density differences at this planar geometry (Figure S1),
which likely explains the energetic differences.
Notably, neither SA-ResHF nor SA-CASSCF captures
the more compact nature of the T state compared to the V state.
SS-CASSCF solutions do capture this distinction\cite{Saade2024JCTC},
indicating that SA-ResHF may not be as efficient at
capturing orbital relaxation as SS-CASSCF.

The most striking feature of Figure \ref{fig:ethene} is
the disappearance of the SS-CASSCF V state at 40 degrees.
Saade and Burton\cite{Saade2024JCTC} attribute this to the
coalescence of two SS-CASSCF solutions, resulting in a
pair annihilation point.
On the other hand, both SA-CASSCF and SA-ResHF produce a continuous V state
energy surface across the full torsional rotation.
The V state is thought to be essential for a prominent
internal conversion pathway of photoexcited ethene to the ground state. \cite{Ben-Nun1998CPL}
The continuity of the excited state surfaces along ethene's torsional rotation
is therefore a significant advantage of state-averaged approaches
in the context of photodynamics applications.

\begin{table*}[htbp]
  \caption{\label{tab:vee}
  Vertical excitation energies (in eV) for selected QUEST database molecules.
  3SA-CASSCF(2e,2o) and 3SA-ResHF(3sd) energies were computed
  using the def2-QZVP basis set.
  $\langle S^2 \rangle$ values for ResHF states are shown in parentheses.
  Theoretical best estimates (TBE)
  obtained from QUEST are used to compute mean average errors (MAE).
  Singlet-triplet energy gaps are calculated as
  $\Delta E_{st} = E_{\ce{S1}} - E_{\ce{T1}}$.}
\begin{ruledtabular}
  \begin{tabular}{lccccccccc}
    & \multicolumn{3}{c}{3SA-CASSCF}
    & \multicolumn{3}{c}{3SA-ResHF}
    & \multicolumn{3}{c}{TBE} \\
   Molecule
   & E$_{\ce{S1}}$ &  E$_{\ce{T1}}$ & $\Delta E_{st}$
   & E$_{\ce{S1}}$ &  E$_{\ce{T1}}$ & $\Delta E_{st}$
   & E$_{\ce{S1}}$ &  E$_{\ce{T1}}$ & $\Delta E_{st}$
   \\ \hline
    formaldehyde &
    3.07 & 2.47 & 0.60 & 
    2.75 (0.064) & 2.40 (2.000) & 0.35 & 
    3.97 & 3.58 & 0.39
    \footnote{Theoretical best estimates (TBE) taken from QUEST subset \#1 (Ref. \onlinecite{QUEST1})}
    \\
    acetaldehyde &
    3.41 & 2.86 & 0.56 & 
    3.12 (0.067) & 2.80 (2.000) & 0.32 & 
    4.31 & 3.98 & 0.33 $^{\text{a}}$ 
    \\
    formamide &
    4.45 & 4.01 & 0.44 & 
    4.38 (0.055) & 4.12 (2.000) & 0.26 & 
    5.63 & 5.37 & 0.26 $^{\text{a}}$ 
    \\
    streptocyanine-C1 &
    8.25 & 5.36 & 2.88 & 
    7.27 (0.093) & 5.70 (2.001) & 1.58 & 
    7.12 & 5.52 & 1.60 $^{\text{a}}$ 
    \\
    acetone &
    3.58 & 3.05 & 0.53 & 
    3.31 (0.066) & 3.01 (2.000) & 0.30 & 
    4.48 & 4.15 & 0.33
    \footnote{Theoretical best estimates (TBE) taken from QUEST subset \#3 (Ref. \onlinecite{QUEST3})}
    \\
    cyclopentadiene &
    6.47 & 3.51 & 2.96 & 
    6.00 (0.027) & 3.37 (2.000) & 2.63 & 
    5.55 & 3.31 & 2.24 $^{\text{b}}$ 
    \\
    pyrimidine &
    5.35 & 4.97 & 0.38 & 
    4.61 (0.454) & 4.43 (2.011) & 0.18 & 
    4.45 & 4.10 & 0.35 $^{\text{b}}$ 
    \\
    benzoxadiazole &
    5.60 & 2.99 & 2.61 & 
    3.95 (0.347) & 3.15 (2.009) & 0.80 & 
    4.52 & 2.74 & 1.78
    \footnote{Theoretical best estimates (TBE) taken from QUEST subset \#7 (Ref. \onlinecite{QUEST7})}
    \\
    benzothiadiazole &
    5.00 & 2.90 & 2.10 & 
    3.69 (0.210) & 3.06 (2.004) & 0.63 & 
    4.23 & 2.82 & 1.41 $^{\text{c}}$ 
   \\ \hline
   MAE &
   0.97 & 0.70 & 0.48 & 
   0.75 & 0.66 & 0.27 & 
  \end{tabular}
\end{ruledtabular}
\end{table*}

\subsection{\label{sec:vee}
High Quality ResHF Singlet-Triplet Gaps}

Accurate singlet-triplet state energy gaps ($\Delta E_{ST}$) are essential
for intersystem crossing.
Even small organic molecules with minimal spin-orbit couplings can exhibit
ultrafast intersystem crossings given small enough $\Delta E_{ST}$.
\cite{Richter2012JPCL, Penfold2012JCP}
To survey the performance of ResHF for $\Delta E_{ST}$,
we compute the lowest lying singlet and triplet excited states for a
set of nine molecules taken from the QUEST
database\cite{QUEST1, QUEST3, QUEST7}.
We apply an equal weight 3SA scheme
across the \ce{S0}, \ce{S1}, and \ce{T1} states
for both CASSCF and ResHF calculations.
To select the initial guess, we first compute configuration
interaction singles calculations of the \ce{S1} state to identify the largest
occupied-to-virtual single excitation contribution from the RHF orbitals.
The subsequent occupied-virtual orbital pair was used to build the
CASSCF and ResHF initial wavefunctions, as shown in Figure \ref{fig:wfn}.

For the ``larger'' molecules (pyrimidine, benzothiadiazole, benzoxadiazole),
ResHF convergence failures
were observed when using the 4 Slater determinant wavefunction.
Steady decreases in energy were observed for the first
10 optimization cycles before reaching a plateau. After about 50 more
cycles, large spikes in energy occur. We hypothesize this is due to the
relatively naive convergence strategy used here (simple DIIS).
Second order, or quasi-second order, convergers are commonly applied to
CASSCF wavefunction optimization, and thus we expect the same
will be required to more reliably converge ResHF for a broader range of systems.

ResHF convergence improved significantly upon omission of the doubly excited
determinant.
Therefore, Table \ref{tab:vee} displays results
from a 3SA-ResHF(3sd) wavefunction.
4SA-ResHF(4sd) energies for the
smaller molecules can be found in the Supplementary Materials.
We emphasize that this reduction of the ResHF wavefunction expansion
is motivated by convergence stability, \emph{not} computational expense.
ResHF scales as $\mathcal{O}(N^2)$ with respect to the number
of Slater determinants in the expansion.

In comparing the vertical excitation energies to the theoretical
best estimates (TBEs), it is clear that both ResHF(3sd) and
CASSCF(2e, 2o) are far from achieving quantitative accuracy
with such a small active space. 3SA-CASSCF shows a 0.97 eV
mean absolute error (MAE) for triplet states and a 0.70 eV
MAE for singlet states. 3SA-ResHF has slightly lower
MAEs, with 0.75 eV for triplet states and 0.66 eV for singlet states.

However, ResHF significantly improves over CASSCF for predicting
$\Delta E_{ST}$, with an MAE of
0.27 eV to CASSCF's 0.48 eV.
This improvement occurs entirely at
the neglect of dynamic correlation.
Qualitative errors in state splittings in CASSCF are well-known.
Conventionally, dynamical correlation
is considered mandatory to reduce systematic errors in
singlet-triplet gaps\cite{Stoneburner2018JCP} and
beyond.\cite{Bettanin2017JCTC, Khokhlov2020JPCA, Segado2016PCCP}
Our results suggest that \emph{relaxation} may be more important
than dynamical correlation in this context. This
fits will with conclusions drawn by Tran and Neuscamman,\cite{Tran2020JPCA}
who showed that SS-CASSCF can dramatically reduce systematic errors
in SA-CASSCF calculations of charge-transfer states.

In examining the ResHF wavefunctions, we commonly find
that each ResHF electronic state is predominantly described by
one or two Slater determinants.
Direct comparison of orbitals between ResHF and CASSCF is not straightforward,
as ResHF has multiple sets of orbitals. We therefore computed
electronic density differences for each \ce{S1} and \ce{T0} state.
Visualizations of these for all molecules in Table \ref{tab:vee}
can be found in the Supplementary Materials.
In all cases, the density differences from SA-CASSCF
and SA-ResHF are qualitatively similar, indicating that
the same electronic states were found by both methods.
However, small deviations in these density differences may
have surprisingly large impacts on the state-splittings.
For instance, in streptocyanine, SA-ResHF underestimates
the $\Delta E_{ST}$ gap by 0.02 eV, while SA-CASSCF overestimates
this same gap by 1.28 eV. Yet, their density differences
appear qualitatively similar (Figure S5).
In another example, signatures of orbital relaxation are evident in
the SA-ResHF density differences
for benzoxadiazole and benzothiadiazole in the form of
a negative band in the plane of the ring (Figures S9 and S10).
Interestingly, while SA-CASSCF overestimates
these $\Delta E_{ST}$ gaps, SA-ResHF underestimates them.

Of course, the current benchmark set
is limited in scope due to limitations in ResHF convergence.
Further investigations on the ability of ResHF to overcome state splitting
errors traditionally attributed to a lack of dynamic
correlation across a wider benchmark set
of molecules is ongoing in our group.

Finally, we note that the largest
errors in $\Delta E_{ST}$ for SA-ResHF occur for
pyrimidine, benzoxadiazole, and benzothiadiazole in which
we also observed notable spin contamination in the singlet states.
See the Supplemental Materials for a derivation of
$\langle S^2\rangle$.
In our view, the increased spin contamination is likely a symptom
pointing to the presence of additional static correlation in these systems
that cannot be captured with a 3 Slater determinant expansion, which may
be mitigated by more robust initial guesses.
Spin contamination can also be entirely eliminated by using a set
of spin-restricted determinants to build the ResHF wavefunction.
Work towards this end is ongoing in our group.

\section{\label{sec:end}
Conclusions and Future Outlook}

In this paper, we have presented a numerically stable reformulation of the ResHF method.
The matrix adjugate ResHF implementation,
enabled by the resolution-of-the-identity approximation,
effectively uses the determinant of
the molecular orbital overlap matrix to remove numerical instabilities encountered
when attempting to take the inverse of a nearly singular overlap matrix.
The results shown here are consistent with
recent efforts by Chen and Scuseria to eliminate numerical instabilities
in the Hartree--Fock--Bogoliubov method. \cite{Chen2023JCP}
Our implementation is freely available in the open-source code, yucca. \cite{yucca}

Using the ResHF-adj implementation, we have benchmarked the
performance of ResHF against CASSCF in several systems.
We showed that ResHF is free from state biasing errors observed in
SA-CASSCF in the avoided crossing arising from LiF bond dissociation.
In fact, the SA-ResHF energy surfaces for the ionic and covalent LiF states
map on directly to the SS-CASSCF energy surfaces.
We also showed that SA-ResHF is able to produce continuous excited state energy
surfaces in the torsional rotation about the carbon--carbon bond of ethene,
in contrast with recently reported SS-CASSCF results.
These two results in particular lead us to conclude that ResHF is able
to unite the energetic benefits of state-specific methods with the
reliability of state-averaged approaches.
We closed with a comparison between ResHF and CASSCF singlet--triplet
energy gaps, showing that a similarly sized ResHF wavefunction can
significantly outperform CASSCF, albeit hindered by spin contamination.

Moving forward, we see several avenues for improvement in the ResHF implementation.
First, the $\mathcal{O}(N^5)$ computational scaling of the matrix adjugate
formulation of ResHF needs to be reduced to make it affordable for
NAMD simulations. We see several routes to reducing the scaling,
including, as an example, a hybrid implementation that uses the
conventional approach for determinant pairs with well-conditioned
overlap matrices, and
the adjugate approach only for pairs with ill-conditioned overlap matrices.
On the other hand, a decomposition of $\xi_{ijk}$ similar to
the tensor train decomposition could also be used to reduce scaling.
Finally, using restricted determinants instead of unrestricted determinants
could also reduce the scaling of the ResHF implementation because
an additional factor of $\sigma_k$ could be factored out of the
$\xi_{ijk}$ term.
A manuscript detailing the use of restricted determinants in ResHF is currently
being prepared.

A more pressing concern is the difficulty of converging ResHF wavefunctions.
Indeed, despite the $\mathcal{O}(N^5)$ scaling of the present implementation,
we find that the convergence of ResHF is the limiting factor in our
current implementation, not the cost.
We see two important routes to improving the convergence of ResHF.
First is to develop a more robust initial guess for the ResHF wavefunction,
which we have observed plays an outsized role in the convergence of ResHF.
Second is to apply second-order optimization techniques.
All of the above-mentioned strategies are under investigation in our group.

There remain several open questions concerning the overall
reliability of the ResHF method for excited states.
The results presented here confirm that dynamical correlation
will be essential to obtaining quantitatively accurate excitation energies
with ResHF. That said, there may still be a role for ResHF in
NAMD simulations even without dynamical correlation, just as
CASSCF is still used in NAMD simulations despite having the
same problem.

In closing, we note that the ResHF method shares many
structural similarities with other nonorthogonal electronic structure
methods, such as breathing orbital valence bond theory\cite{Hiberty2002TCA}
and multistate density functional theory. \cite{Lu2022JPCL}
Therefore, we hope the matrix adjugate approach to ResHF
will help to push forward the practicality of these under-developed
and newly-developed methods.

\section*{Supplemental Material}
The Supplemental Material includes 4SA-ResHF(4sd) results for
singlet-triplet gaps, a derivation of $\langle\hat{S}^2\rangle$
for nonorthogonal Slater determinants,
and plots of density differences.

\begin{acknowledgments}
  This work was supported by a startup fund from Case Western Reserve University
  and NSF CAREER award CHE-2236959.
  This work made use of the High Performance Computing Resource in the Core
  Facility for Advanced Research Computing at Case Western Reserve University.
  E.R.M. was supported by the Molecular Sciences Software Institute
  under U.S. National Science Foundation Grant No. ACI-1547580.
\end{acknowledgments}


\section*{Data Availability}
The ResHF implementation used in this study is openly available
at \href{https://gitlab.com/team-parker/yucca}{https://gitlab.com/team-parker/yucca}.
The code and data that support the findings of this study
are openly available in ``Numerically Stable Resonating Hartree-Fock
with Matrix Adjugates'' at
\href{https://osf.io/ahnbp/}{https://osf.io/ahnbp/}.


\section*{Author Contributions}
\textbf{Ericka Roy Miller:}
    conceptualization (supporting);
    data curation (lead);
    investigation (lead);
    formal analysis (lead);
    methodology (equal);
    software (lead);
    visualization (lead);
    writing - original draft (lead);
    writing - review and editing (equal).
    \textbf{Shane M. Parker:}
    conceptualization (lead);
    funding acquisition (lead);
    formal analysis (supporting);
    methodology (equal);
    resources (lead);
    software (supporting);
    supervision (lead);
    writing - review and editing (equal).


\appendix

\section{ \label{sec:df}
  Applying Resolution-of-the-Identity to Resonating Fock Matrices}

As detailed in Section \ref{sec:res-adj}, the bulk of the complexity in
ResHF-adj is contained in the virtual, occupied block of
the $\mathbf{K}^{AB}$ matrix. Therefore, we will focus exclusively on
Eq. \eqref{eq:kvo} in the following analysis.
In our implementation, we generally separate out $\mathbf{K}^{AB}$
into four terms:
\begin{subequations}
\begin{align}
  \mathbf{K}^{AB}_{\text{uo}} &=
  \mathbf{A}_{\text{u}}^\dagger
  (\mathbf{K1} + \mathbf{K2} + \mathbf{K3} + \mathbf{K4})
  \mathbf{A}_{\text{o}} \\
  \mathbf{K1} &= s_{AB} \mathbf{F}^{AB} ~ \mathbf{Q}^{AB} ~ \mathbf{S} \\
  \mathbf{K2} &= -s_{AB} \mathbf{S} ~ \mathbf{Q}^{AB}
  \mathbf{F}^{AB} ~ \mathbf{Q}^{AB} ~ \mathbf{S} \\
  \mathbf{K3} &= H_{AB} \mathbf{S} ~ \mathbf{Q}^{AB} ~ \mathbf{S} \\
  \mathbf{K4} &= - s_{AB} E_I \mathbf{S} ~ \mathbf{Q}^{AB} ~ \mathbf{S} .
\end{align}
\end{subequations}
For the sake of organization, we will apply this same grouping throughout
our discussion of the resolution-of-the-identity (RI) algorithms.

To streamline the equations in this section, we introduce
Einstein notation and re-write Eq. \eqref{eq:df} as
\begin{equation}
  (p q | r s) \approx (pq|\tilde{P})  (\tilde{P}|rs),
\end{equation}
where
\begin{equation}
  (pq|\tilde{P}) = (p q | Q ) (L^{-1})_{QP}
\end{equation}
and $L_{PQ}$ comes from the Cholesky decomposition of the $(P|Q)$ matrix.

We collect here all
the two electron components of Eq. \eqref{eq:kvo} expressed using
unrestricted spin MOs (see Appendix \ref{sec:ureshf}) in the
biorthogonal SVD MO representation:
\begin{subequations} \label{eq:ukvo2e}
\begin{align}
  (K^{AB, \alpha})^{2e-}_{ai} &=
  (K1^\alpha)^{2e-}_{ai} + (K2^\alpha)^{2e-}_{ai} + (K3^\alpha)^{2e-}_{ai}
  \\
  (K1^\alpha)^{2e-}_{ai} &= \label{eq:K1}
  \eta (a_A^\alpha i_B^\alpha | j_A^\beta j_B^\beta)
  \xi_{i}^\alpha \xi_{j}^\beta
  \nonumber \\
  &+ \eta \determ{\boldsymbol{\Sigma}^{AB,\beta}}
  (a_A^\alpha i_B^\alpha | | j_A^\alpha j_B^\alpha)
  \xi_{ij}^\alpha
  \\
  (K2^\alpha)^{2e-}_{ai} &= \label{eq:K2}
  - \eta \bar{S}^{AB, \alpha}_{aj}
  (j_A^\alpha i_B^\alpha | k_A^\beta k_B^\beta)
  \xi_{ij}^\alpha \xi_{k}^\beta
  \nonumber \\
  &- \eta \determ{\boldsymbol{\Sigma}^{AB,\beta}}
  \bar{S}^{AB,\alpha}_{aj}
  (j_A^\alpha i_B^\alpha | | k_A^\alpha k_B^\alpha)
  \xi_{ijk}^\alpha
  \\
  (K3^\alpha)^{2e-}_{ai} &= \label{eq:K3}
  \eta
  \bar{S}^{AB,\alpha}_{ai}
  (j_A^\alpha j_B^\alpha | k_A^\beta k_B^\beta)
  \xi_{ij}^\alpha \xi_{k}^\beta
  \nonumber \\
  &+ \frac{1}{2} \eta
  \bar{S}^{AB,\alpha}_{ai}
  (j_A^\beta j_B^\beta | | k_A^\beta k_B^\beta)
  \xi_{i}^\alpha \xi_{jk}^\beta
  \\
  &+ \frac{1}{2} \eta \determ{\boldsymbol{\Sigma}^{AB,\beta}}
  \bar{S}^{AB,\alpha}_{ai}
  (j_A^\alpha j_B^\alpha | | k_A^\alpha k_B^\alpha)
  \xi_{ijk}^\alpha
  ~ . \nonumber
\end{align}
\end{subequations}
We now present the RI algorithms used to compute each set of
$(K^{AB, \alpha})^{2e-}_{ai}$ terms.
For simplicity, we use $T$ to denote a generic intermediate, and we
do not highlight the use of common intermediates across different contributions.
Furthermore, we show only the $\alpha$ component. The $\beta$ component
can be obtained by swapping $\alpha$ and $\beta$ indices.

\subsection{RI approximation of two-electron $\mathbf{K}^{AB}_{\text{uo}}$ Set 1 terms}

Expressing the ERIs using auxiliary basis functions,
we write Eq. \eqref{eq:K1} as
\begin{align}
  (K1^\alpha)^{2e-}_{ai} &=
    \eta (a_A^\alpha i_B^\alpha | \tilde{P}) (\tilde{P} | j_A^\beta j_B^\beta)
    \xi_{i}^\alpha \xi_{j}^\beta
    \nonumber \\
    &+ \eta \determ{\boldsymbol{\Sigma}^{AB,\beta}}
    (a_A^\alpha i_B^\alpha | \tilde{P}) (\tilde{P} | j_A^\alpha j_B^\alpha)
    \xi_{ij}^\alpha
    \\
    &- \eta \determ{\boldsymbol{\Sigma}^{AB,\beta}}
    (a_A^\alpha j_B^\alpha | \tilde{P}) (\tilde{P} | j_A^\alpha i_B^\alpha)
    \xi_{ij}^\alpha
    ~ . \nonumber
\end{align}
The first $(K1^\alpha)^{2e-}_{ai}$ term, the mixed-spin Coulomb contribution, is
fully contracted over the spin MO indices prior to the final contraction step:
\begin{subequations}
\begin{align}
  T_{aiP} &\leftarrow \xi_{i}^\alpha (a_A^\alpha i_B^\alpha | \tilde{P}) \\
  T_{P} &\leftarrow (\tilde{P} | j_A^\beta j_B^\beta) \xi_{j}^\beta  \\
  (K1^\alpha)^{2e-}_{ai} &\mathrel{+}= \eta T_{aiP} T_{P} .
\end{align}
\end{subequations}
For the second $(K1^\alpha)^{2e-}_{ai}$ term,
the spin-pure Coulomb contribution, we are also able to fully
contract over the spin MO indices in the intermediate step:
\begin{subequations}
\begin{align}
  T_{Pi} &\leftarrow (\tilde{P} | j_A^\alpha j_B^\alpha) \xi_{ij}^\alpha \\
  (K1^\alpha)^{2e-}_{ai} &\mathrel{+}= \eta (a_A^\alpha i_B^\alpha | \tilde{P}) T_{Pi}.  \\
\end{align}
\end{subequations}
For the final $(K1^\alpha)^{2e-}_{ai}$ term,
the exchange contribution, we are unable to fully contract
over the spin MO indices in the intermediate step.
Thus, the final contraction is
between two Rank 3 tensors:
\begin{subequations}
\begin{align}
  T_{Pji} &\leftarrow (\tilde{P} | j_A^\alpha i_B^\alpha) \xi_{ij}^\alpha \\
  (K1^\alpha)^{2e-}_{ai} &\mathrel{-}= \eta (a_A^\alpha j_B^\alpha | \tilde{P}) T_{Pji}
\end{align}
\end{subequations}

\subsection{RI approximation of two-electron $\mathbf{K}^{AB}_{\text{uo}}$ Set 2 terms}

Expressing the ERIs using auxiliary basis functions,
we write Eq. \eqref{eq:K2} as
\begin{align}
  (K2^\alpha)^{2e-}_{ai} &=
    - \eta \bar{S}^{AB,\alpha}_{aj}
    (j_A^\alpha i_B^\alpha | \tilde{P}) (\tilde{P} | k_A^\beta k_B^\beta)
    \xi_{ij}^\alpha \xi_{k}^\beta
    \\
    &- \eta \determ{\boldsymbol{\Sigma}^{AB,\beta}}
    \bar{S}^{AB,\alpha}_{aj}
    (j_A^\alpha i_B^\alpha | \tilde{P}) (\tilde{P} | k_A^\alpha k_B^\alpha)
    \xi_{ijk}^\alpha
    \nonumber \\
    &+ \eta \determ{\boldsymbol{\Sigma}^{AB,\beta}}
    \bar{S}^{AB,\alpha}_{aj}
    (j_A^\alpha k_B^\alpha | \tilde{P}) (\tilde{P} | k_A^\alpha i_B^\alpha)
    \xi_{ijk}^\alpha
    . \nonumber
\end{align}
To compute the first $(K2^\alpha)^{2e-}_{ai}$ term,
the mixed-spin Coulomb contribution, we first combine the
overlap and $\xi_{ij}$ terms,
before proceeding with contractions involving the ERIs:
\begin{subequations}
\begin{align}
  T_{aji} &\leftarrow \bar{S}^{AB,\alpha}_{aj} \xi_{ij}^\alpha \\
  T_{aiP} &\leftarrow T_{aji} (j_A^\alpha i_B^\alpha | \tilde{P}) \\
  T_{P} &\leftarrow (\tilde{P}  | k_A^\beta k_B^\beta) \xi_{k}^\beta \\
  (K2^\alpha)^{2e-}_{ai} &\mathrel{-}= \eta T_{aiP} T_{P}
  .
\end{align}
\end{subequations}
For the second $(K2^\alpha)^{2e-}_{ai}$ term,
the spin-pure Coulomb contribution, we
incorporate the overlap term via a Rank 4 tensor,
before completing the tensor contraction with the other half ERI:
\begin{subequations}
\begin{align}
  T_{ajiP} &\leftarrow \bar{S}^{AB,\alpha}_{aj}(j_A^\alpha i_B^\alpha | \tilde{P}) \\
  T_{Pij} &\leftarrow (\tilde{P} | k_A^\alpha k_B^\alpha) \xi_{ijk}^{\alpha} \\
  (K2^\alpha)^{2e-}_{ai} &\mathrel{-}= \eta \determ{\boldsymbol{\Sigma}^{AB,\beta}} T_{ajiP} T_{Pij}
  .
\end{align}
\end{subequations}
Finally, in the third $(K2^\alpha)^{2e-}_{ai}$ term,
the exchange contribution is formed through the contraction
of two Rank 4 tensors,
\begin{subequations}
\begin{align}
  T_{ajkP} &\leftarrow \bar{S}^{AB,\alpha}_{aj} (j_A^\alpha k_B^\alpha | \tilde{P}) \\
  T_{Pkij} &\leftarrow (\tilde{P} | k_A^\alpha i_B^\alpha) \xi_{ijk}^\alpha \\
  (K2^\alpha)^{2e-}_{ai} &\mathrel{+}= \eta \determ{\boldsymbol{\Sigma}^{AB,\beta}} T_{ajkP} T_{Pkij}
  ,
\end{align}
\end{subequations}
making this the largest scaling step in Resonating Fock matrix builder routine
at O($N^5$) with respect to the number of basis functions.

\subsection{RI approximation of two-electron $\mathbf{K}^{AB}_{\text{uo}}$ Set 3 terms}

Expressing the ERIs using auxiliary basis functions,
we write Eq. \eqref{eq:K3} as
\begin{align}
  (K3^\alpha)^{2e-}_{ai} &=
    \eta \bar{S}^{AB,\alpha}_{ai}
    (j_A^\alpha j_B^\alpha | \tilde{P}) (\tilde{P} | k_A^\beta k_B^\beta)
    \xi_{ij}^\alpha \xi_{k}^\beta
    \\
    &+ \frac{1}{2} \eta
    \bar{S}^{AB, \alpha}_{ai} \xi_{i}^\alpha
    (j_A^\beta j_B^\beta | \tilde{P}) (\tilde{P} | k_A^\beta k_B^\beta)
    \xi_{jk}^\beta
    \nonumber \\
    &- \frac{1}{2} \eta
    \bar{S}^{AB,\alpha}_{ai} \xi_{i}^\alpha
    (j_A^\beta k_B^\beta | \tilde{P}) (\tilde{P} | k_A^\beta j_B^\beta)
    \xi_{jk}^\beta
    \nonumber \\
    &+ \frac{1}{2} \eta \determ{\boldsymbol{\Sigma}^{AB,\beta}}
    \bar{S}^{AB, \alpha}_{ai}
    (j_A^\alpha j_B^\alpha | \tilde{P}) (\tilde{P} | k_A^\alpha k_B^\alpha)
    \xi_{ijk}^\alpha
    \nonumber \\
    &- \frac{1}{2} \eta \determ{\boldsymbol{\Sigma}^{AB,\beta}}
    \bar{S}^{AB, \alpha}_{ai}
    (j_A^\alpha k_B^\alpha | \tilde{P}) (\tilde{P} | k_A^\alpha j_B^\alpha)
    \xi_{ijk}^\alpha
    .\nonumber
\end{align}
For the first term in $(K3^\alpha)^{2e-}_{ai}$,
the mixed-spin Coulomb contribution, we are able to fully
contract over the spin MO indices before the final contraction step:
\begin{subequations}
\begin{align}
  T_{iP} &\leftarrow \xi_{ij}^\alpha (j_A^\alpha j_B^\alpha | \tilde{P}) \\
  T_{P} &\leftarrow (\tilde{P} | k_A^\beta k_B^\beta) \xi_{k}^\beta \\
  (K3^\alpha)^{2e-}_{ai} &\mathrel{+}= \eta \bar{S}^{AB,\alpha}_{ai} T_{iP} T_{P}
  .
\end{align}
\end{subequations}
For the second term in $(K3^\alpha)^{2e-}_{ai}$,
the $\beta$ spin Coulomb contribution, we contract the half ERIs
into Rank 2 tensors before fully contracting with the $\xi_{jk}$ term to produce a
scalar, which is then used to scale the final operation:
\begin{subequations}
\begin{align}
  T_{jP} &\leftarrow (j_A^\alpha j_B^\alpha |  \tilde{P}) \\
  T_{Pk} &\leftarrow (\tilde{P}  | k_A^\alpha k_B^\alpha) \\
  T_{jk} &\leftarrow T_{jP} T_{Pk}  \\
  T &\leftarrow T_{jk} \xi_{jk}^\beta \\
  (K3^\alpha)^{2e-}_{ai} &\mathrel{+}= \frac{1}{2} \eta
  \bar{S}^{AB, \alpha}_{ai} \xi_{i}^\alpha T
  .
\end{align}
\end{subequations}
We apply a similar strategy for the third term in $(K3^\alpha)^{2e-}_{ai}$,
the $\beta$ spin exchange contribution,
albeit contracting Rank 3 tensors to produce the scalar,
which is then used in the final operation:
\begin{subequations}
\begin{align}
  T_{Pkj} &\leftarrow (\tilde{P} | k_A^\beta j_B^\beta) \xi_{jk}^\beta \\
  T &\leftarrow (j_A^\beta k_B^\beta | \tilde{P}) T_{Pkj} \\
  (K3^\alpha)^{2e-}_{ai} &\mathrel{-}= \frac{1}{2} \eta
  \bar{S}^{AB, \alpha}_{ai} \xi_{i}^\alpha T
  .
\end{align}
\end{subequations}
For the fourth term in $(K3^\alpha)^{2e-}_{ai}$, the
$\alpha$ spin Coulomb contribution, we again contract the half ERIs
into Rank 2 tensors
before contracting with the $\xi_{ijk}$ and overlap terms sequentially:
\begin{subequations}
\begin{align}
  T_{jP} &\leftarrow (j_A^\alpha j_B^\alpha | \tilde{P}) \\
  T_{Pk} &\leftarrow (\tilde{P} | k_A^\alpha k_B^\alpha) \\
  T_{jk} &\leftarrow T_{jP} T_{Pk} \\
  T_i &\leftarrow \xi_{ijk}^\alpha T_{jk} \\
  (K3^\alpha)^{2e-}_{ai} &\mathrel{+}=
  \frac{1}{2} \eta \determ{\boldsymbol{\Sigma}^{AB,\beta}}
  \bar{S}^{AB, \alpha}_{ai} T_i
  .
\end{align}
\end{subequations}
Finally, to resolve the fifth term in $(K3^\alpha)^{2e-}_{ai}$,
the $\alpha$ spin exchange contribution,
we contract the ERIs via a Rank 4 tensor and a Rank 3 tensor
before the final contraction with the overlap term:
\begin{subequations}
\begin{align}
  T_{Pijk} &\leftarrow (\tilde{P}  | k_A^\alpha j_B^\alpha) \xi_{ijk}^\alpha \\
  T_i &\leftarrow (j_A^\alpha k_B^\alpha | \tilde{P}) T_{Pijk} \\
  (K3^\alpha)^{2e-}_{ai} &\mathrel{-}=
  \frac{1}{2} \eta \determ{\boldsymbol{\Sigma}^{AB,\beta}}
  \bar{S}^{AB, \alpha}_{ai} T_i
  .
\end{align}
\end{subequations}

\section{ \label{sec:ureshf}
  ResHF intermediates with unrestricted orbitals}

In the following treatment, we will express equations using unrestricted
molecular orbitals,
\begin{gather}
  \phi^{A,\omega}_p (\mathbf{x_1}) = s(\omega) \psi^{A,\omega}_p (\mathbf{r_1})
\end{gather}
where we have defined a generic spin index, $\omega = \alpha, \beta$,
and $\psi^{A,\omega}_p$ is a spatial MOs
corresponding to $\omega$ spin electrons.
The MO overlap matrix of Eq. \eqref{eq:Smo} becomes a diagonal block matrix
with elements
\begin{align} \label{eq:uSmo}
  \mathbf{S}^{AB,\omega}
  = \mathbf{C}^{A,\omega,\dagger} \mathbf{S} \mathbf{C}^{B,\omega},
\end{align}
and the interdeterminant overlap can be separated into
$\alpha$ and $\beta$ components as
\begin{equation} \label{eq:udetsab}
  s_{AB}
  = \determ{\mathbf{S}^{AB, \alpha}_{\text{oo}}} \determ{\mathbf{S}^{AB, \beta}_{\text{oo}}} \\
  =  \eta  \left[ \prod^{N_\alpha}_i \sigma^\alpha_i \right]
  \left[ \prod^{N_\beta}_j \sigma^\beta_j \right] ,
\end{equation}
where two sets of singular values, $\sigma^\omega_i$, are distinguished
based on spin and $\eta = \determ{\mathbf{U}^{AB,\alpha} (\mathbf{V}^{AB,\alpha})^{\dagger}}
\determ{\mathbf{U}^{AB,\beta} (\mathbf{V}^{AB,\beta})^{\dagger}} = \pm 1$.

The interdeterminant density matrix of Eq. \eqref{eq:density} also becomes
block diagonal with respect to spin, with elements
\begin{equation} \label{eq:ugamma}
  \boldsymbol{\gamma}^{AB,\omega} = s_{AB}  \mathbf{Q}^{AB,\omega}.
\end{equation}
Using this, we can write the coupling matrix element from
Eq. \eqref{eq:hab} as
\begin{align} \label{eq:uhab}
  H_{AB} =
  s_{AB} \sum_\omega \left[ \trace{\mathbf{h}  \mathbf{Q}^{AB,\omega} }
+ \frac{1}{2}
  \trace{\mathbf{G}^{AB,\omega}  \mathbf{Q}^{AB,\omega}} \right]
  ,
\end{align}
where
\begin{align} \label{eq:ugab}
  G^{AB, \omega}_{\mu\nu} =
  \sum_{\lambda\sigma} \left( \mu \nu | | \lambda \sigma \right)
  {Q}^{AB, \omega}_{\sigma\lambda}
  + \sum_{\lambda\sigma} \left( \mu \nu | \lambda \sigma \right)
  {Q}^{AB, \omega'}_{\sigma\lambda} ,
\end{align}
where $\omega \neq \omega'$
and the electron repulsion integrals (ERIs) are now spatial integrals,
defined as
\begin{equation}
  (\mu \nu | \lambda \sigma) =
  \iint d\mathbf{r}_1 \mathbf{r}_2
  \chi^*_\mu (\mathbf{r}_1) \chi_\nu (\mathbf{r}_1) \hat{r}^{-1}_{1 2}
  \chi^*_\lambda (\mathbf{r}_2) \chi_\sigma (\mathbf{r}_2) .
\end{equation}
We can also render the AO basis $\mathbf{K}^{AB}$ matrix from
Eq. \eqref{eq:kabao} into a block diagonal matrix with elements
\begin{multline}
  \mathbf{K}^{AB,\omega} =
  s_{AB}
  (\mathbf{1} - \mathbf{S} ~ \mathbf{Q}^{AB,\omega}) ~
  \mathbf{F}^{AB,\omega} ~ \mathbf{Q}^{AB,\omega} ~ \mathbf{S} \\
  + (H_{AB} - s_{AB} E_I) ~
  \mathbf{S} ~ \mathbf{Q}^{AB,\omega} ~ \mathbf{S} ,
\end{multline}
where $\mathbf{F}^{AB,\omega} = \mathbf{h} + \mathbf{G}^{AB,\omega}$.

\bibliography{reshf}

\begin{thebibliography}{56}%
\makeatletter
\providecommand \@ifxundefined [1]{%
 \@ifx{#1\undefined}
}%
\providecommand \@ifnum [1]{%
 \ifnum #1\expandafter \@firstoftwo
 \else \expandafter \@secondoftwo
 \fi
}%
\providecommand \@ifx [1]{%
 \ifx #1\expandafter \@firstoftwo
 \else \expandafter \@secondoftwo
 \fi
}%
\providecommand \natexlab [1]{#1}%
\providecommand \enquote  [1]{``#1''}%
\providecommand \bibnamefont  [1]{#1}%
\providecommand \bibfnamefont [1]{#1}%
\providecommand \citenamefont [1]{#1}%
\providecommand \href@noop [0]{\@secondoftwo}%
\providecommand \href [0]{\begingroup \@sanitize@url \@href}%
\providecommand \@href[1]{\@@startlink{#1}\@@href}%
\providecommand \@@href[1]{\endgroup#1\@@endlink}%
\providecommand \@sanitize@url [0]{\catcode `\\12\catcode `\$12\catcode
  `\&12\catcode `\#12\catcode `\^12\catcode `\_12\catcode `\%12\relax}%
\providecommand \@@startlink[1]{}%
\providecommand \@@endlink[0]{}%
\providecommand \url  [0]{\begingroup\@sanitize@url \@url }%
\providecommand \@url [1]{\endgroup\@href {#1}{\urlprefix }}%
\providecommand \urlprefix  [0]{URL }%
\providecommand \Eprint [0]{\href }%
\providecommand \doibase [0]{http://dx.doi.org/}%
\providecommand \selectlanguage [0]{\@gobble}%
\providecommand \bibinfo  [0]{\@secondoftwo}%
\providecommand \bibfield  [0]{\@secondoftwo}%
\providecommand \translation [1]{[#1]}%
\providecommand \BibitemOpen [0]{}%
\providecommand \bibitemStop [0]{}%
\providecommand \bibitemNoStop [0]{.\EOS\space}%
\providecommand \EOS [0]{\spacefactor3000\relax}%
\providecommand \BibitemShut  [1]{\csname bibitem#1\endcsname}%
\let\auto@bib@innerbib\@empty
\bibitem [{\citenamefont {Richter}\ \emph {et~al.}(2012)\citenamefont
  {Richter}, \citenamefont {Marquetand}, \citenamefont {González-Vázquez},
  \citenamefont {Sola},\ and\ \citenamefont {González}}]{Richter2012JPCL}%
  \BibitemOpen
  \bibfield  {author} {\bibinfo {author} {\bibfnamefont {M.}~\bibnamefont
  {Richter}}, \bibinfo {author} {\bibfnamefont {P.}~\bibnamefont {Marquetand}},
  \bibinfo {author} {\bibfnamefont {J.}~\bibnamefont {González-Vázquez}},
  \bibinfo {author} {\bibfnamefont {I.}~\bibnamefont {Sola}}, \ and\ \bibinfo
  {author} {\bibfnamefont {L.}~\bibnamefont {González}},\ }\href {\doibase
  10.1021/jz301312h} {\bibfield  {journal} {\bibinfo  {journal} {J. Phys. Chem.
  Lett.}\ }\textbf {\bibinfo {volume} {3}},\ \bibinfo {pages} {3090} (\bibinfo
  {year} {2012})}\BibitemShut {NoStop}%
\bibitem [{\citenamefont {Penfold}\ \emph {et~al.}(2012)\citenamefont
  {Penfold}, \citenamefont {Spesyvtsev}, \citenamefont {Kirkby}, \citenamefont
  {Minns}, \citenamefont {Parker}, \citenamefont {Fielding},\ and\
  \citenamefont {Worth}}]{Penfold2012JCP}%
  \BibitemOpen
  \bibfield  {author} {\bibinfo {author} {\bibfnamefont {T.}~\bibnamefont
  {Penfold}}, \bibinfo {author} {\bibfnamefont {R.}~\bibnamefont {Spesyvtsev}},
  \bibinfo {author} {\bibfnamefont {O.~M.}\ \bibnamefont {Kirkby}}, \bibinfo
  {author} {\bibfnamefont {R.~S.}\ \bibnamefont {Minns}}, \bibinfo {author}
  {\bibfnamefont {D.~S.~N.}\ \bibnamefont {Parker}}, \bibinfo {author}
  {\bibfnamefont {H.~H.}\ \bibnamefont {Fielding}}, \ and\ \bibinfo {author}
  {\bibfnamefont {G.~A.}\ \bibnamefont {Worth}},\ }\href {\doibase
  10.1063/1.4767054} {\bibfield  {journal} {\bibinfo  {journal} {J. Chem.
  Phys.}\ }\textbf {\bibinfo {volume} {137}},\ \bibinfo {pages} {204310}
  (\bibinfo {year} {2012})}\BibitemShut {NoStop}%
\bibitem [{\citenamefont {Curchod}, \citenamefont {Sisto},\ and\ \citenamefont
  {Martínez}(2017)}]{Curchod2017JCPA}%
  \BibitemOpen
  \bibfield  {author} {\bibinfo {author} {\bibfnamefont {B.~F.~E.}\
  \bibnamefont {Curchod}}, \bibinfo {author} {\bibfnamefont {A.}~\bibnamefont
  {Sisto}}, \ and\ \bibinfo {author} {\bibfnamefont {T.~J.}\ \bibnamefont
  {Martínez}},\ }\href {\doibase 10.1021/acs.jpca.6b09962} {\bibfield
  {journal} {\bibinfo  {journal} {J. Phys. Chem. A}\ }\textbf {\bibinfo
  {volume} {121}},\ \bibinfo {pages} {265} (\bibinfo {year}
  {2017})}\BibitemShut {NoStop}%
\bibitem [{\citenamefont {Nelson}\ \emph {et~al.}(2014)\citenamefont {Nelson},
  \citenamefont {Fernandez-Alberti}, \citenamefont {Roitberg},\ and\
  \citenamefont {Tretiak}}]{Nelson2014ACR}%
  \BibitemOpen
  \bibfield  {author} {\bibinfo {author} {\bibfnamefont {T.}~\bibnamefont
  {Nelson}}, \bibinfo {author} {\bibfnamefont {S.}~\bibnamefont
  {Fernandez-Alberti}}, \bibinfo {author} {\bibfnamefont {A.~E.}\ \bibnamefont
  {Roitberg}}, \ and\ \bibinfo {author} {\bibfnamefont {S.}~\bibnamefont
  {Tretiak}},\ }\href {\doibase 10.1021/ar400263p} {\bibfield  {journal}
  {\bibinfo  {journal} {Acc. Chem. Res.}\ }\textbf {\bibinfo {volume} {47}},\
  \bibinfo {pages} {1155} (\bibinfo {year} {2014})}\BibitemShut {NoStop}%
\bibitem [{\citenamefont {Borne}\ \emph {et~al.}(2024)\citenamefont {Borne},
  \citenamefont {Cooper}, \citenamefont {Ashfold}, \citenamefont {Bachmann},
  \citenamefont {Bhattacharyya}, \citenamefont {Boll}, \citenamefont
  {Bonanomi}, \citenamefont {Bosch}, \citenamefont {Callegari}, \citenamefont
  {Centurion}, \citenamefont {Coreno}, \citenamefont {Curchod}, \citenamefont
  {Danailov}, \citenamefont {Demidovich}, \citenamefont {Di~Fraia},
  \citenamefont {Erk}, \citenamefont {Faccialà}, \citenamefont {Feifel},
  \citenamefont {Forbes}, \citenamefont {Hansen}, \citenamefont {Holland},
  \citenamefont {Ingle}, \citenamefont {Lindh}, \citenamefont {Ma},
  \citenamefont {McGhee}, \citenamefont {Muvva}, \citenamefont {Nunes},
  \citenamefont {Odate}, \citenamefont {Pathak}, \citenamefont {Plekan},
  \citenamefont {Prince}, \citenamefont {Rebernik}, \citenamefont {Rouzée},
  \citenamefont {Rudenko}, \citenamefont {Simoncig}, \citenamefont {Squibb},
  \citenamefont {Venkatachalam}, \citenamefont {Vozzi}, \citenamefont {Weber},
  \citenamefont {Kirrander},\ and\ \citenamefont {Rolles}}]{Borne2024NC}%
  \BibitemOpen
  \bibfield  {author} {\bibinfo {author} {\bibfnamefont {K.~D.}\ \bibnamefont
  {Borne}}, \bibinfo {author} {\bibfnamefont {J.~C.}\ \bibnamefont {Cooper}},
  \bibinfo {author} {\bibfnamefont {M.~N.~R.}\ \bibnamefont {Ashfold}},
  \bibinfo {author} {\bibfnamefont {J.}~\bibnamefont {Bachmann}}, \bibinfo
  {author} {\bibfnamefont {S.}~\bibnamefont {Bhattacharyya}}, \bibinfo {author}
  {\bibfnamefont {R.}~\bibnamefont {Boll}}, \bibinfo {author} {\bibfnamefont
  {M.}~\bibnamefont {Bonanomi}}, \bibinfo {author} {\bibfnamefont
  {M.}~\bibnamefont {Bosch}}, \bibinfo {author} {\bibfnamefont
  {C.}~\bibnamefont {Callegari}}, \bibinfo {author} {\bibfnamefont
  {M.}~\bibnamefont {Centurion}}, \bibinfo {author} {\bibfnamefont
  {M.}~\bibnamefont {Coreno}}, \bibinfo {author} {\bibfnamefont {B.~F.~E.}\
  \bibnamefont {Curchod}}, \bibinfo {author} {\bibfnamefont {M.~B.}\
  \bibnamefont {Danailov}}, \bibinfo {author} {\bibfnamefont {A.}~\bibnamefont
  {Demidovich}}, \bibinfo {author} {\bibfnamefont {M.}~\bibnamefont
  {Di~Fraia}}, \bibinfo {author} {\bibfnamefont {B.}~\bibnamefont {Erk}},
  \bibinfo {author} {\bibfnamefont {D.}~\bibnamefont {Faccialà}}, \bibinfo
  {author} {\bibfnamefont {R.}~\bibnamefont {Feifel}}, \bibinfo {author}
  {\bibfnamefont {R.~J.~G.}\ \bibnamefont {Forbes}}, \bibinfo {author}
  {\bibfnamefont {C.~S.}\ \bibnamefont {Hansen}}, \bibinfo {author}
  {\bibfnamefont {D.~M.~P.}\ \bibnamefont {Holland}}, \bibinfo {author}
  {\bibfnamefont {R.~A.}\ \bibnamefont {Ingle}}, \bibinfo {author}
  {\bibfnamefont {R.}~\bibnamefont {Lindh}}, \bibinfo {author} {\bibfnamefont
  {L.}~\bibnamefont {Ma}}, \bibinfo {author} {\bibfnamefont {H.~G.}\
  \bibnamefont {McGhee}}, \bibinfo {author} {\bibfnamefont {S.~B.}\
  \bibnamefont {Muvva}}, \bibinfo {author} {\bibfnamefont {J.~P.~F.}\
  \bibnamefont {Nunes}}, \bibinfo {author} {\bibfnamefont {A.}~\bibnamefont
  {Odate}}, \bibinfo {author} {\bibfnamefont {S.}~\bibnamefont {Pathak}},
  \bibinfo {author} {\bibfnamefont {O.}~\bibnamefont {Plekan}}, \bibinfo
  {author} {\bibfnamefont {K.~C.}\ \bibnamefont {Prince}}, \bibinfo {author}
  {\bibfnamefont {P.}~\bibnamefont {Rebernik}}, \bibinfo {author}
  {\bibfnamefont {A.}~\bibnamefont {Rouzée}}, \bibinfo {author} {\bibfnamefont
  {A.}~\bibnamefont {Rudenko}}, \bibinfo {author} {\bibfnamefont
  {A.}~\bibnamefont {Simoncig}}, \bibinfo {author} {\bibfnamefont {R.~J.}\
  \bibnamefont {Squibb}}, \bibinfo {author} {\bibfnamefont {A.~S.}\
  \bibnamefont {Venkatachalam}}, \bibinfo {author} {\bibfnamefont
  {C.}~\bibnamefont {Vozzi}}, \bibinfo {author} {\bibfnamefont {P.~M.}\
  \bibnamefont {Weber}}, \bibinfo {author} {\bibfnamefont {A.}~\bibnamefont
  {Kirrander}}, \ and\ \bibinfo {author} {\bibfnamefont {D.}~\bibnamefont
  {Rolles}},\ }\href {\doibase 10.1038/s41557-023-01420-w} {\bibfield
  {journal} {\bibinfo  {journal} {Nat. Chem.}\ }\textbf {\bibinfo {volume}
  {16}},\ \bibinfo {pages} {499} (\bibinfo {year} {2024})}\BibitemShut
  {NoStop}%
\bibitem [{\citenamefont {Schuurman}\ and\ \citenamefont
  {Stolow}(2018)}]{Schuurman2018ARPC}%
  \BibitemOpen
  \bibfield  {author} {\bibinfo {author} {\bibfnamefont {M.~S.}\ \bibnamefont
  {Schuurman}}\ and\ \bibinfo {author} {\bibfnamefont {A.}~\bibnamefont
  {Stolow}},\ }\href {\doibase 10.1146/annurev-physchem-052516-050721}
  {\bibfield  {journal} {\bibinfo  {journal} {Annu. Rev. Phys. Chem.}\ }\textbf
  {\bibinfo {volume} {69}},\ \bibinfo {pages} {427} (\bibinfo {year}
  {2018})}\BibitemShut {NoStop}%
\bibitem [{\citenamefont {Janoš}\ and\ \citenamefont
  {Slavíček}(2023)}]{Janos2023JCTC}%
  \BibitemOpen
  \bibfield  {author} {\bibinfo {author} {\bibfnamefont {J.}~\bibnamefont
  {Janoš}}\ and\ \bibinfo {author} {\bibfnamefont {P.}~\bibnamefont
  {Slavíček}},\ }\href {\doibase 10.1021/acs.jctc.3c00908} {\bibfield
  {journal} {\bibinfo  {journal} {J. Chem. Theory Comput.}\ }\textbf {\bibinfo
  {volume} {19}},\ \bibinfo {pages} {8273} (\bibinfo {year}
  {2023})}\BibitemShut {NoStop}%
\bibitem [{\citenamefont {Papineau}, \citenamefont {Jacquemin},\ and\
  \citenamefont {Vacher}(2024)}]{Papineau2024JPCL}%
  \BibitemOpen
  \bibfield  {author} {\bibinfo {author} {\bibfnamefont {T.~V.}\ \bibnamefont
  {Papineau}}, \bibinfo {author} {\bibfnamefont {D.}~\bibnamefont {Jacquemin}},
  \ and\ \bibinfo {author} {\bibfnamefont {M.}~\bibnamefont {Vacher}},\ }\href
  {\doibase 10.1021/acs.jpclett.3c03014} {\bibfield  {journal} {\bibinfo
  {journal} {J. Phys. Chem. Lett.}\ }\textbf {\bibinfo {volume} {15}},\
  \bibinfo {pages} {636} (\bibinfo {year} {2024})}\BibitemShut {NoStop}%
\bibitem [{\citenamefont {Mukherjee}\ \emph {et~al.}(2024)\citenamefont
  {Mukherjee}, \citenamefont {Mattos}, \citenamefont {Toldo}, \citenamefont
  {Lischka},\ and\ \citenamefont {Barbatti}}]{Mukherjee2024JCP}%
  \BibitemOpen
  \bibfield  {author} {\bibinfo {author} {\bibfnamefont {S.}~\bibnamefont
  {Mukherjee}}, \bibinfo {author} {\bibfnamefont {R.~S.}\ \bibnamefont
  {Mattos}}, \bibinfo {author} {\bibfnamefont {J.~M.}\ \bibnamefont {Toldo}},
  \bibinfo {author} {\bibfnamefont {H.}~\bibnamefont {Lischka}}, \ and\
  \bibinfo {author} {\bibfnamefont {M.}~\bibnamefont {Barbatti}},\ }\href
  {\doibase 10.1063/5.0203636} {\bibfield  {journal} {\bibinfo  {journal} {J.
  Chem. Phys.}\ }\textbf {\bibinfo {volume} {160}},\ \bibinfo {pages} {154306}
  (\bibinfo {year} {2024})}\BibitemShut {NoStop}%
\bibitem [{\citenamefont {Levine}\ \emph {et~al.}(2006)\citenamefont {Levine},
  \citenamefont {Ko}, \citenamefont {Quenneville},\ and\ \citenamefont
  {Martínez}}]{Levine2006MP}%
  \BibitemOpen
  \bibfield  {author} {\bibinfo {author} {\bibfnamefont {B.~G.}\ \bibnamefont
  {Levine}}, \bibinfo {author} {\bibfnamefont {C.}~\bibnamefont {Ko}}, \bibinfo
  {author} {\bibfnamefont {J.}~\bibnamefont {Quenneville}}, \ and\ \bibinfo
  {author} {\bibfnamefont {T.~J.}\ \bibnamefont {Martínez}},\ }\href {\doibase
  10.1080/00268970500417762} {\bibfield  {journal} {\bibinfo  {journal} {Mol.
  Phys.}\ }\textbf {\bibinfo {volume} {104}},\ \bibinfo {pages} {1039}
  (\bibinfo {year} {2006})}\BibitemShut {NoStop}%
\bibitem [{\citenamefont {Taylor}, \citenamefont {Tozer},\ and\ \citenamefont
  {Curchod}(2023)}]{Taylor2023JCP}%
  \BibitemOpen
  \bibfield  {author} {\bibinfo {author} {\bibfnamefont {J.~T.}\ \bibnamefont
  {Taylor}}, \bibinfo {author} {\bibfnamefont {D.~J.}\ \bibnamefont {Tozer}}, \
  and\ \bibinfo {author} {\bibfnamefont {B.~F.~E.}\ \bibnamefont {Curchod}},\
  }\href {\doibase 10.1063/5.0176140} {\bibfield  {journal} {\bibinfo
  {journal} {J. Chem. Phys.}\ }\textbf {\bibinfo {volume} {159}},\ \bibinfo
  {pages} {214115} (\bibinfo {year} {2023})}\BibitemShut {NoStop}%
\bibitem [{\citenamefont {Taylor}, \citenamefont {Tozer},\ and\ \citenamefont
  {Curchod}(2024)}]{Taylor2024JPCA}%
  \BibitemOpen
  \bibfield  {author} {\bibinfo {author} {\bibfnamefont {J.~T.}\ \bibnamefont
  {Taylor}}, \bibinfo {author} {\bibfnamefont {D.~J.}\ \bibnamefont {Tozer}}, \
  and\ \bibinfo {author} {\bibfnamefont {B.~F.~E.}\ \bibnamefont {Curchod}},\
  }\href {\doibase 10.1021/acs.jpca.4c02503} {\bibfield  {journal} {\bibinfo
  {journal} {J. Phys. Chem. A}\ }\textbf {\bibinfo {volume} {128}},\ \bibinfo
  {pages} {5314} (\bibinfo {year} {2024})}\BibitemShut {NoStop}%
\bibitem [{\citenamefont {Segado}, \citenamefont {Gómez},\ and\ \citenamefont
  {Reguero}(2016)}]{Segado2016PCCP}%
  \BibitemOpen
  \bibfield  {author} {\bibinfo {author} {\bibfnamefont {M.}~\bibnamefont
  {Segado}}, \bibinfo {author} {\bibfnamefont {I.}~\bibnamefont {Gómez}}, \
  and\ \bibinfo {author} {\bibfnamefont {M.}~\bibnamefont {Reguero}},\ }\href
  {\doibase 10.1039/C5CP04690D} {\bibfield  {journal} {\bibinfo  {journal}
  {Phys. Chem. Chem. Phys.}\ }\textbf {\bibinfo {volume} {18}},\ \bibinfo
  {pages} {6861} (\bibinfo {year} {2016})}\BibitemShut {NoStop}%
\bibitem [{\citenamefont {Tran}\ and\ \citenamefont
  {Neuscamman}(2020)}]{Tran2020JPCA}%
  \BibitemOpen
  \bibfield  {author} {\bibinfo {author} {\bibfnamefont {L.~N.}\ \bibnamefont
  {Tran}}\ and\ \bibinfo {author} {\bibfnamefont {E.}~\bibnamefont
  {Neuscamman}},\ }\href {\doibase 10.1021/acs.jpca.0c07593} {\bibfield
  {journal} {\bibinfo  {journal} {J. Phys. Chem. A}\ }\textbf {\bibinfo
  {volume} {124}},\ \bibinfo {pages} {8273} (\bibinfo {year}
  {2020})}\BibitemShut {NoStop}%
\bibitem [{\citenamefont {Marie}\ and\ \citenamefont
  {Burton}(2023)}]{Marie2023JPCA}%
  \BibitemOpen
  \bibfield  {author} {\bibinfo {author} {\bibfnamefont {A.}~\bibnamefont
  {Marie}}\ and\ \bibinfo {author} {\bibfnamefont {H.~G.~A.}\ \bibnamefont
  {Burton}},\ }\href {\doibase 10.1021/acs.jpca.3c00603} {\bibfield  {journal}
  {\bibinfo  {journal} {J. Phys. Chem. A}\ }\textbf {\bibinfo {volume} {127}},\
  \bibinfo {pages} {4538} (\bibinfo {year} {2023})}\BibitemShut {NoStop}%
\bibitem [{\citenamefont {Saade}\ and\ \citenamefont
  {Burton}(2024)}]{Saade2024JCTC}%
  \BibitemOpen
  \bibfield  {author} {\bibinfo {author} {\bibfnamefont {S.}~\bibnamefont
  {Saade}}\ and\ \bibinfo {author} {\bibfnamefont {H.~G.~A.}\ \bibnamefont
  {Burton}},\ }\href {\doibase 10.1021/acs.jctc.4c00212} {\bibfield  {journal}
  {\bibinfo  {journal} {J. Chem. Theory Comput.}\ }\textbf {\bibinfo {volume}
  {20}},\ \bibinfo {pages} {5105} (\bibinfo {year} {2024})}\BibitemShut
  {NoStop}%
\bibitem [{\citenamefont {Bremond}(1964)}]{Bremond1964NP}%
  \BibitemOpen
  \bibfield  {author} {\bibinfo {author} {\bibfnamefont {B.}~\bibnamefont
  {Bremond}},\ }\href {\doibase 10.1016/0029-5582(64)90579-6} {\bibfield
  {journal} {\bibinfo  {journal} {Nuc. Phys.}\ }\textbf {\bibinfo {volume}
  {58}},\ \bibinfo {pages} {687} (\bibinfo {year} {1964})}\BibitemShut
  {NoStop}%
\bibitem [{\citenamefont {Fukutome}(1988)}]{Fukutome1988PoTP}%
  \BibitemOpen
  \bibfield  {author} {\bibinfo {author} {\bibfnamefont {H.}~\bibnamefont
  {Fukutome}},\ }\href {\doibase 10.1143/PTP.80.417} {\bibfield  {journal}
  {\bibinfo  {journal} {Prog. Theor. Phys.}\ }\textbf {\bibinfo {volume}
  {80}},\ \bibinfo {pages} {417} (\bibinfo {year} {1988})}\BibitemShut
  {NoStop}%
\bibitem [{\citenamefont {Burton}\ and\ \citenamefont
  {Thom}(2019)}]{Burton2019JCTC}%
  \BibitemOpen
  \bibfield  {author} {\bibinfo {author} {\bibfnamefont {H.~G.~A.}\
  \bibnamefont {Burton}}\ and\ \bibinfo {author} {\bibfnamefont {A.~J.~W.}\
  \bibnamefont {Thom}},\ }\href {\doibase 10.1021/acs.jctc.9b00441} {\bibfield
  {journal} {\bibinfo  {journal} {J. Chem. Theory Comput.}\ }\textbf {\bibinfo
  {volume} {15}},\ \bibinfo {pages} {4851} (\bibinfo {year}
  {2019})}\BibitemShut {NoStop}%
\bibitem [{\citenamefont {Hiberty}\ and\ \citenamefont
  {Shaik}(2002)}]{Hiberty2002TCA}%
  \BibitemOpen
  \bibfield  {author} {\bibinfo {author} {\bibfnamefont {P.~C.}\ \bibnamefont
  {Hiberty}}\ and\ \bibinfo {author} {\bibfnamefont {S.}~\bibnamefont
  {Shaik}},\ }\href {\doibase 10.1007/s00214-002-0364-8} {\bibfield  {journal}
  {\bibinfo  {journal} {Theor. Chem. Acc.}\ }\textbf {\bibinfo {volume}
  {108}},\ \bibinfo {pages} {255} (\bibinfo {year} {2002})}\BibitemShut
  {NoStop}%
\bibitem [{\citenamefont {Lu}\ and\ \citenamefont {Gao}(2022)}]{Lu2022JPCL}%
  \BibitemOpen
  \bibfield  {author} {\bibinfo {author} {\bibfnamefont {Y.}~\bibnamefont
  {Lu}}\ and\ \bibinfo {author} {\bibfnamefont {J.}~\bibnamefont {Gao}},\
  }\href {\doibase 10.1021/acs.jpclett.2c02088} {\bibfield  {journal} {\bibinfo
   {journal} {J. Phys. Chem. Lett.}\ }\textbf {\bibinfo {volume} {13}},\
  \bibinfo {pages} {7762} (\bibinfo {year} {2022})}\BibitemShut {NoStop}%
\bibitem [{\citenamefont {Olsen}(2015)}]{Olsen2015JCP}%
  \BibitemOpen
  \bibfield  {author} {\bibinfo {author} {\bibfnamefont {J.}~\bibnamefont
  {Olsen}},\ }\href {\doibase 10.1063/1.4929724} {\bibfield  {journal}
  {\bibinfo  {journal} {J. Chem. Phys.}\ }\textbf {\bibinfo {volume} {143}},\
  \bibinfo {pages} {114102} (\bibinfo {year} {2015})}\BibitemShut {NoStop}%
\bibitem [{\citenamefont {Sun}, \citenamefont {Gao},\ and\ \citenamefont
  {Scuseria}(2024)}]{Sun2024JCTC}%
  \BibitemOpen
  \bibfield  {author} {\bibinfo {author} {\bibfnamefont {C.}~\bibnamefont
  {Sun}}, \bibinfo {author} {\bibfnamefont {F.}~\bibnamefont {Gao}}, \ and\
  \bibinfo {author} {\bibfnamefont {G.~E.}\ \bibnamefont {Scuseria}},\ }\href
  {\doibase 10.1021/acs.jctc.4c00240} {\bibfield  {journal} {\bibinfo
  {journal} {J. Chem. Theory Comput.}\ }\textbf {\bibinfo {volume} {20}},\
  \bibinfo {pages} {3741} (\bibinfo {year} {2024})}\BibitemShut {NoStop}%
\bibitem [{\citenamefont {Jiménez-Hoyos}, \citenamefont {Rodríguez-Guzmán},\
  and\ \citenamefont {Scuseria}(2013)}]{Jimenez-Hoyos2013JCP}%
  \BibitemOpen
  \bibfield  {author} {\bibinfo {author} {\bibfnamefont {C.~A.}\ \bibnamefont
  {Jiménez-Hoyos}}, \bibinfo {author} {\bibfnamefont {R.}~\bibnamefont
  {Rodríguez-Guzmán}}, \ and\ \bibinfo {author} {\bibfnamefont {G.~E.}\
  \bibnamefont {Scuseria}},\ }\href {\doibase 10.1063/1.4840097} {\bibfield
  {journal} {\bibinfo  {journal} {J. Chem. Phys.}\ }\textbf {\bibinfo {volume}
  {139}},\ \bibinfo {pages} {224110} (\bibinfo {year} {2013})}\BibitemShut
  {NoStop}%
\bibitem [{\citenamefont {Nite}\ and\ \citenamefont
  {{Jim{\'e}nez-Hoyos}}(2019)}]{Nite2019JCTC}%
  \BibitemOpen
  \bibfield  {author} {\bibinfo {author} {\bibfnamefont {J.}~\bibnamefont
  {Nite}}\ and\ \bibinfo {author} {\bibfnamefont {C.~A.}\ \bibnamefont
  {{Jim{\'e}nez-Hoyos}}},\ }\href {\doibase 10.1021/acs.jctc.9b00579}
  {\bibfield  {journal} {\bibinfo  {journal} {J. Chem. Theory Comput.}\
  }\textbf {\bibinfo {volume} {15}},\ \bibinfo {pages} {5343} (\bibinfo {year}
  {2019})}\BibitemShut {NoStop}%
\bibitem [{\citenamefont {Mahler}\ and\ \citenamefont
  {Thompson}(2021)}]{Mahler2021JCP}%
  \BibitemOpen
  \bibfield  {author} {\bibinfo {author} {\bibfnamefont {A.~D.}\ \bibnamefont
  {Mahler}}\ and\ \bibinfo {author} {\bibfnamefont {L.~M.}\ \bibnamefont
  {Thompson}},\ }\href {\doibase 10.1063/5.0053615} {\bibfield  {journal}
  {\bibinfo  {journal} {J. Chem. Phys.}\ }\textbf {\bibinfo {volume} {154}},\
  \bibinfo {pages} {244101} (\bibinfo {year} {2021})}\BibitemShut {NoStop}%
\bibitem [{\citenamefont {Hill}\ and\ \citenamefont
  {Underwood}(1985)}]{Hill1985SJoAaDM}%
  \BibitemOpen
  \bibfield  {author} {\bibinfo {author} {\bibfnamefont {R.~D.}\ \bibnamefont
  {Hill}}\ and\ \bibinfo {author} {\bibfnamefont {E.~E.}\ \bibnamefont
  {Underwood}},\ }\href {\doibase 10.1137/0606071} {\bibfield  {journal}
  {\bibinfo  {journal} {SIAM. J. on Algebraic and Discrete Methods}\ }\textbf
  {\bibinfo {volume} {6}},\ \bibinfo {pages} {731} (\bibinfo {year}
  {1985})}\BibitemShut {NoStop}%
\bibitem [{\citenamefont {Stewart}(1998)}]{Stewart1998LAaiA}%
  \BibitemOpen
  \bibfield  {author} {\bibinfo {author} {\bibfnamefont {G.~W.}\ \bibnamefont
  {Stewart}},\ }\href {\doibase 10.1016/S0024-3795(98)10098-8} {\bibfield
  {journal} {\bibinfo  {journal} {Linear Algebra and its Applications}\
  }\textbf {\bibinfo {volume} {283}},\ \bibinfo {pages} {151} (\bibinfo {year}
  {1998})}\BibitemShut {NoStop}%
\bibitem [{\citenamefont {Koch}\ and\ \citenamefont
  {Dalgaard}(1993)}]{Koch1993CPL}%
  \BibitemOpen
  \bibfield  {author} {\bibinfo {author} {\bibfnamefont {H.}~\bibnamefont
  {Koch}}\ and\ \bibinfo {author} {\bibfnamefont {E.}~\bibnamefont
  {Dalgaard}},\ }\href {\doibase 10.1016/0009-2614(93)87129-Q} {\bibfield
  {journal} {\bibinfo  {journal} {Chem. Phys. Lett.}\ }\textbf {\bibinfo
  {volume} {212}},\ \bibinfo {pages} {193} (\bibinfo {year}
  {1993})}\BibitemShut {NoStop}%
\bibitem [{\citenamefont {Chen}\ and\ \citenamefont
  {Scuseria}(2023)}]{Chen2023JCP}%
  \BibitemOpen
  \bibfield  {author} {\bibinfo {author} {\bibfnamefont {G.~P.}\ \bibnamefont
  {Chen}}\ and\ \bibinfo {author} {\bibfnamefont {G.~E.}\ \bibnamefont
  {Scuseria}},\ }\href {\doibase 10.1063/5.0156124} {\bibfield  {journal}
  {\bibinfo  {journal} {J. Chem. Phys.}\ }\textbf {\bibinfo {volume} {158}},\
  \bibinfo {pages} {231102} (\bibinfo {year} {2023})}\BibitemShut {NoStop}%
\bibitem [{\citenamefont {Parker}(2024)}]{yucca}%
  \BibitemOpen
  \bibfield  {author} {\bibinfo {author} {\bibfnamefont {S.~M.}\ \bibnamefont
  {Parker}},\ }\href@noop {} {\enquote {\bibinfo {title} {Yucca},}\ }\bibinfo
  {howpublished} {\url{https://gitlab.com/team-parker/yucca}} (\bibinfo {year}
  {2024})\BibitemShut {NoStop}%
\bibitem [{\citenamefont {Burton}(2021)}]{Burton2021JCP}%
  \BibitemOpen
  \bibfield  {author} {\bibinfo {author} {\bibfnamefont {H.~G.~A.}\
  \bibnamefont {Burton}},\ }\href {\doibase 10.1063/5.0045442} {\bibfield
  {journal} {\bibinfo  {journal} {J. Chem. Phys.}\ }\textbf {\bibinfo {volume}
  {154}},\ \bibinfo {pages} {144109} (\bibinfo {year} {2021})}\BibitemShut
  {NoStop}%
\bibitem [{\citenamefont {Broer}\ and\ \citenamefont
  {Nieuwpoort}(1988)}]{Broer1988TCA}%
  \BibitemOpen
  \bibfield  {author} {\bibinfo {author} {\bibfnamefont {R.}~\bibnamefont
  {Broer}}\ and\ \bibinfo {author} {\bibfnamefont {W.~C.}\ \bibnamefont
  {Nieuwpoort}},\ }\href {\doibase 10.1007/BF00527744} {\bibfield  {journal}
  {\bibinfo  {journal} {Theoret. Chim. Acta}\ }\textbf {\bibinfo {volume}
  {73}},\ \bibinfo {pages} {405} (\bibinfo {year} {1988})}\BibitemShut
  {NoStop}%
\bibitem [{\citenamefont {Baerends}, \citenamefont {Ellis},\ and\ \citenamefont
  {Ros}(1973)}]{Baerends1973CP}%
  \BibitemOpen
  \bibfield  {author} {\bibinfo {author} {\bibfnamefont {E.~J.}\ \bibnamefont
  {Baerends}}, \bibinfo {author} {\bibfnamefont {D.~E.}\ \bibnamefont {Ellis}},
  \ and\ \bibinfo {author} {\bibfnamefont {P.}~\bibnamefont {Ros}},\
  }\href@noop {} {\bibfield  {journal} {\bibinfo  {journal} {Chem. Phys.}\
  }\textbf {\bibinfo {volume} {2}},\ \bibinfo {pages} {41} (\bibinfo {year}
  {1973})}\BibitemShut {NoStop}%
\bibitem [{\citenamefont {Dunlap}, \citenamefont {Connolly},\ and\
  \citenamefont {Sabin}(1979)}]{Dunlap79JChemPhys}%
  \BibitemOpen
  \bibfield  {author} {\bibinfo {author} {\bibfnamefont {B.~I.}\ \bibnamefont
  {Dunlap}}, \bibinfo {author} {\bibfnamefont {J.~W.~D.}\ \bibnamefont
  {Connolly}}, \ and\ \bibinfo {author} {\bibfnamefont {J.~R.}\ \bibnamefont
  {Sabin}},\ }\href@noop {} {\bibfield  {journal} {\bibinfo  {journal} {J.
  Chem. Phys.}\ }\textbf {\bibinfo {volume} {71}},\ \bibinfo {pages} {3396}
  (\bibinfo {year} {1979})}\BibitemShut {NoStop}%
\bibitem [{\citenamefont {Weigend}(2002)}]{Weigend2002PCCP}%
  \BibitemOpen
  \bibfield  {author} {\bibinfo {author} {\bibfnamefont {F.}~\bibnamefont
  {Weigend}},\ }\href {\doibase 10.1039/B204199P} {\bibfield  {journal}
  {\bibinfo  {journal} {Phys. Chem. Chem. Phys.}\ }\textbf {\bibinfo {volume}
  {4}},\ \bibinfo {pages} {4285} (\bibinfo {year} {2002})}\BibitemShut
  {NoStop}%
\bibitem [{\citenamefont {Neese}(2022)}]{Neese2022WIREs}%
  \BibitemOpen
  \bibfield  {author} {\bibinfo {author} {\bibfnamefont {F.}~\bibnamefont
  {Neese}},\ }\href {\doibase 10.1002/wcms.1606} {\bibfield  {journal}
  {\bibinfo  {journal} {WIREs Comput. Mol. Sci.}\ }\textbf {\bibinfo {volume}
  {12}},\ \bibinfo {pages} {e1606} (\bibinfo {year} {2022})}\BibitemShut
  {NoStop}%
\bibitem [{\citenamefont {Miller}\ and\ \citenamefont {Parker}(2024)}]{OSFadj}%
  \BibitemOpen
  \bibfield  {author} {\bibinfo {author} {\bibfnamefont {E.~R.}\ \bibnamefont
  {Miller}}\ and\ \bibinfo {author} {\bibfnamefont {S.~M.}\ \bibnamefont
  {Parker}},\ }\href@noop {} {\enquote {\bibinfo {title} {Numerically stable
  resonating hartree-fock with matrix adjugates},}\ }\bibinfo {howpublished}
  {\url{https://osf.io/ahnbp/}} (\bibinfo {year} {2024})\BibitemShut {NoStop}%
\bibitem [{\citenamefont {Kollmar}\ \emph {et~al.}(2019)\citenamefont
  {Kollmar}, \citenamefont {Sivalingam}, \citenamefont {Helmich-Paris},
  \citenamefont {Angeli},\ and\ \citenamefont {Neese}}]{Kollmar2019JCC}%
  \BibitemOpen
  \bibfield  {author} {\bibinfo {author} {\bibfnamefont {C.}~\bibnamefont
  {Kollmar}}, \bibinfo {author} {\bibfnamefont {K.}~\bibnamefont {Sivalingam}},
  \bibinfo {author} {\bibfnamefont {B.}~\bibnamefont {Helmich-Paris}}, \bibinfo
  {author} {\bibfnamefont {C.}~\bibnamefont {Angeli}}, \ and\ \bibinfo {author}
  {\bibfnamefont {F.}~\bibnamefont {Neese}},\ }\href {\doibase
  10.1002/jcc.25801} {\bibfield  {journal} {\bibinfo  {journal} {J. Comp.
  Chem.}\ }\textbf {\bibinfo {volume} {40}},\ \bibinfo {pages} {1463} (\bibinfo
  {year} {2019})}\BibitemShut {NoStop}%
\bibitem [{\citenamefont {Pulay}(1982)}]{Pulay1982JCC}%
  \BibitemOpen
  \bibfield  {author} {\bibinfo {author} {\bibfnamefont {P.}~\bibnamefont
  {Pulay}},\ }\href {\doibase 10.1002/jcc.540030413} {\bibfield  {journal}
  {\bibinfo  {journal} {J. Comp. Chem.}\ }\textbf {\bibinfo {volume} {3}},\
  \bibinfo {pages} {556} (\bibinfo {year} {1982})}\BibitemShut {NoStop}%
\bibitem [{\citenamefont {Weigend}\ and\ \citenamefont
  {Ahlrichs}(2005)}]{Weigend2005PCCP}%
  \BibitemOpen
  \bibfield  {author} {\bibinfo {author} {\bibfnamefont {F.}~\bibnamefont
  {Weigend}}\ and\ \bibinfo {author} {\bibfnamefont {R.}~\bibnamefont
  {Ahlrichs}},\ }\href {\doibase 10.1039/B508541A} {\bibfield  {journal}
  {\bibinfo  {journal} {Phys. Chem. Chem. Phys.}\ }\textbf {\bibinfo {volume}
  {7}},\ \bibinfo {pages} {3297} (\bibinfo {year} {2005})}\BibitemShut
  {NoStop}%
\bibitem [{\citenamefont {Dunning}(1989)}]{Dunning1989JCP}%
  \BibitemOpen
  \bibfield  {author} {\bibinfo {author} {\bibfnamefont {T.~H.}\ \bibnamefont
  {Dunning}},\ }\href {\doibase 10.1063/1.456153} {\bibfield  {journal}
  {\bibinfo  {journal} {J. Chem. Phys.}\ }\textbf {\bibinfo {volume} {90}},\
  \bibinfo {pages} {1007} (\bibinfo {year} {1989})}\BibitemShut {NoStop}%
\bibitem [{\citenamefont {Kendall}, \citenamefont {Dunning},\ and\
  \citenamefont {Harrison}(1992)}]{Kendall1992JCP}%
  \BibitemOpen
  \bibfield  {author} {\bibinfo {author} {\bibfnamefont {R.~A.}\ \bibnamefont
  {Kendall}}, \bibinfo {author} {\bibfnamefont {T.~H.}\ \bibnamefont
  {Dunning}}, \ and\ \bibinfo {author} {\bibfnamefont {R.~J.}\ \bibnamefont
  {Harrison}},\ }\href {\doibase 10.1063/1.462569} {\bibfield  {journal}
  {\bibinfo  {journal} {J. Chem. Phys.}\ }\textbf {\bibinfo {volume} {96}},\
  \bibinfo {pages} {6796} (\bibinfo {year} {1992})}\BibitemShut {NoStop}%
\bibitem [{\citenamefont {Weigend}, \citenamefont {Furche},\ and\ \citenamefont
  {Ahlrichs}(2003)}]{Weigend2003JCP}%
  \BibitemOpen
  \bibfield  {author} {\bibinfo {author} {\bibfnamefont {F.}~\bibnamefont
  {Weigend}}, \bibinfo {author} {\bibfnamefont {F.}~\bibnamefont {Furche}}, \
  and\ \bibinfo {author} {\bibfnamefont {R.}~\bibnamefont {Ahlrichs}},\ }\href
  {\doibase 10.1063/1.1627293} {\bibfield  {journal} {\bibinfo  {journal} {J.
  Chem. Phys.}\ }\textbf {\bibinfo {volume} {119}},\ \bibinfo {pages} {12753}
  (\bibinfo {year} {2003})}\BibitemShut {NoStop}%
\bibitem [{\citenamefont {Malmqvist}\ and\ \citenamefont
  {Roos}(1989)}]{Malmqvist1989CPL}%
  \BibitemOpen
  \bibfield  {author} {\bibinfo {author} {\bibfnamefont {P.-A.}\ \bibnamefont
  {Malmqvist}}\ and\ \bibinfo {author} {\bibfnamefont {B.~O.}\ \bibnamefont
  {Roos}},\ }\href {\doibase 10.1016/0009-2614(89)85347-3} {\bibfield
  {journal} {\bibinfo  {journal} {Chem. Phys. Lett.}\ }\textbf {\bibinfo
  {volume} {155}},\ \bibinfo {pages} {189} (\bibinfo {year}
  {1989})}\BibitemShut {NoStop}%
\bibitem [{\citenamefont {Kahn}, \citenamefont {Hay},\ and\ \citenamefont
  {Shavitt}(1974)}]{Kahn1974JCP}%
  \BibitemOpen
  \bibfield  {author} {\bibinfo {author} {\bibfnamefont {L.~R.}\ \bibnamefont
  {Kahn}}, \bibinfo {author} {\bibfnamefont {P.~J.}\ \bibnamefont {Hay}}, \
  and\ \bibinfo {author} {\bibfnamefont {I.}~\bibnamefont {Shavitt}},\ }\href
  {\doibase 10.1063/1.1682533} {\bibfield  {journal} {\bibinfo  {journal} {J.
  Chem. Phys.}\ }\textbf {\bibinfo {volume} {61}},\ \bibinfo {pages} {3530}
  (\bibinfo {year} {1974})}\BibitemShut {NoStop}%
\bibitem [{\citenamefont {Bauschlicher}\ and\ \citenamefont
  {Langhoff}(1988)}]{Bauschlicher1988JCP}%
  \BibitemOpen
  \bibfield  {author} {\bibinfo {author} {\bibfnamefont {J.}~\bibnamefont
  {Bauschlicher}, \bibfnamefont {Charles~W.}}\ and\ \bibinfo {author}
  {\bibfnamefont {S.~R.}\ \bibnamefont {Langhoff}},\ }\href {\doibase
  10.1063/1.455702} {\bibfield  {journal} {\bibinfo  {journal} {J. Chem.
  Phys.}\ }\textbf {\bibinfo {volume} {89}},\ \bibinfo {pages} {4246} (\bibinfo
  {year} {1988})}\BibitemShut {NoStop}%
\bibitem [{\citenamefont {Ben-Nun}\ and\ \citenamefont
  {Martínez}(2000)}]{Ben-Nun2000CP}%
  \BibitemOpen
  \bibfield  {author} {\bibinfo {author} {\bibfnamefont {M.}~\bibnamefont
  {Ben-Nun}}\ and\ \bibinfo {author} {\bibfnamefont {T.~J.}\ \bibnamefont
  {Martínez}},\ }\href {\doibase 10.1016/S0301-0104(00)00194-4} {\bibfield
  {journal} {\bibinfo  {journal} {Chem. Phys.}\ }\textbf {\bibinfo {volume}
  {259}},\ \bibinfo {pages} {237} (\bibinfo {year} {2000})}\BibitemShut
  {NoStop}%
\bibitem [{\citenamefont {Feller}, \citenamefont {Peterson},\ and\
  \citenamefont {Davidson}(2014)}]{Feller2014JCP}%
  \BibitemOpen
  \bibfield  {author} {\bibinfo {author} {\bibfnamefont {D.}~\bibnamefont
  {Feller}}, \bibinfo {author} {\bibfnamefont {K.~A.}\ \bibnamefont
  {Peterson}}, \ and\ \bibinfo {author} {\bibfnamefont {E.~R.}\ \bibnamefont
  {Davidson}},\ }\href {\doibase 10.1063/1.4894482} {\bibfield  {journal}
  {\bibinfo  {journal} {J. Chem. Phys.}\ }\textbf {\bibinfo {volume} {141}},\
  \bibinfo {pages} {104302} (\bibinfo {year} {2014})}\BibitemShut {NoStop}%
\bibitem [{\citenamefont {Ben-Nun}\ and\ \citenamefont
  {Martínez}(1998)}]{Ben-Nun1998CPL}%
  \BibitemOpen
  \bibfield  {author} {\bibinfo {author} {\bibfnamefont {M.}~\bibnamefont
  {Ben-Nun}}\ and\ \bibinfo {author} {\bibfnamefont {T.~J.}\ \bibnamefont
  {Martínez}},\ }\href {\doibase 10.1016/S0009-2614(98)01115-4} {\bibfield
  {journal} {\bibinfo  {journal} {Chem. Phys. Lett.}\ }\textbf {\bibinfo
  {volume} {298}},\ \bibinfo {pages} {57} (\bibinfo {year} {1998})}\BibitemShut
  {NoStop}%
\bibitem [{\citenamefont {Loos}\ \emph {et~al.}(2018)\citenamefont {Loos},
  \citenamefont {Scemama}, \citenamefont {Blondel}, \citenamefont {Garniron},
  \citenamefont {Caffarel},\ and\ \citenamefont {Jacquemin}}]{QUEST1}%
  \BibitemOpen
  \bibfield  {author} {\bibinfo {author} {\bibfnamefont {P.-F.}\ \bibnamefont
  {Loos}}, \bibinfo {author} {\bibfnamefont {A.}~\bibnamefont {Scemama}},
  \bibinfo {author} {\bibfnamefont {A.}~\bibnamefont {Blondel}}, \bibinfo
  {author} {\bibfnamefont {Y.}~\bibnamefont {Garniron}}, \bibinfo {author}
  {\bibfnamefont {M.}~\bibnamefont {Caffarel}}, \ and\ \bibinfo {author}
  {\bibfnamefont {D.}~\bibnamefont {Jacquemin}},\ }\href {\doibase
  10.1021/acs.jctc.8b00406} {\bibfield  {journal} {\bibinfo  {journal} {J.
  Chem. Theory Comput.}\ }\textbf {\bibinfo {volume} {14}},\ \bibinfo {pages}
  {4360} (\bibinfo {year} {2018})}\BibitemShut {NoStop}%
\bibitem [{\citenamefont {Loos}\ \emph {et~al.}(2020)\citenamefont {Loos},
  \citenamefont {Lipparini}, \citenamefont {Boggio-Pasqua}, \citenamefont
  {Scemama},\ and\ \citenamefont {Jacquemin}}]{QUEST3}%
  \BibitemOpen
  \bibfield  {author} {\bibinfo {author} {\bibfnamefont {P.-F.}\ \bibnamefont
  {Loos}}, \bibinfo {author} {\bibfnamefont {F.}~\bibnamefont {Lipparini}},
  \bibinfo {author} {\bibfnamefont {M.}~\bibnamefont {Boggio-Pasqua}}, \bibinfo
  {author} {\bibfnamefont {A.}~\bibnamefont {Scemama}}, \ and\ \bibinfo
  {author} {\bibfnamefont {D.}~\bibnamefont {Jacquemin}},\ }\href {\doibase
  10.1021/acs.jctc.9b01216} {\bibfield  {journal} {\bibinfo  {journal} {J.
  Chem. Theory Comput.}\ }\textbf {\bibinfo {volume} {16}},\ \bibinfo {pages}
  {1711} (\bibinfo {year} {2020})}\BibitemShut {NoStop}%
\bibitem [{\citenamefont {Loos}\ and\ \citenamefont
  {Jacquemin}(2021)}]{QUEST7}%
  \BibitemOpen
  \bibfield  {author} {\bibinfo {author} {\bibfnamefont {P.-F.}\ \bibnamefont
  {Loos}}\ and\ \bibinfo {author} {\bibfnamefont {D.}~\bibnamefont
  {Jacquemin}},\ }\href {\doibase 10.1021/acs.jpca.1c08524} {\bibfield
  {journal} {\bibinfo  {journal} {J. Phys. Chem. A}\ }\textbf {\bibinfo
  {volume} {125}},\ \bibinfo {pages} {10174} (\bibinfo {year}
  {2021})}\BibitemShut {NoStop}%
\bibitem [{\citenamefont {Stoneburner}, \citenamefont {Truhlar},\ and\
  \citenamefont {Gagliardi}(2018)}]{Stoneburner2018JCP}%
  \BibitemOpen
  \bibfield  {author} {\bibinfo {author} {\bibfnamefont {S.~J.}\ \bibnamefont
  {Stoneburner}}, \bibinfo {author} {\bibfnamefont {D.~G.}\ \bibnamefont
  {Truhlar}}, \ and\ \bibinfo {author} {\bibfnamefont {L.}~\bibnamefont
  {Gagliardi}},\ }\href {\doibase 10.1063/1.5017132} {\bibfield  {journal}
  {\bibinfo  {journal} {J. Chem. Phys.}\ }\textbf {\bibinfo {volume} {148}},\
  \bibinfo {pages} {064108} (\bibinfo {year} {2018})}\BibitemShut {NoStop}%
\bibitem [{\citenamefont {Bettanin}\ \emph {et~al.}(2017)\citenamefont
  {Bettanin}, \citenamefont {Ferrão}, \citenamefont {Pinheiro}, \citenamefont
  {Aquino}, \citenamefont {Lischka}, \citenamefont {Machado},\ and\
  \citenamefont {Nachtigallova}}]{Bettanin2017JCTC}%
  \BibitemOpen
  \bibfield  {author} {\bibinfo {author} {\bibfnamefont {F.}~\bibnamefont
  {Bettanin}}, \bibinfo {author} {\bibfnamefont {L.~F.~A.}\ \bibnamefont
  {Ferrão}}, \bibinfo {author} {\bibfnamefont {M.~J.}\ \bibnamefont
  {Pinheiro}}, \bibinfo {author} {\bibfnamefont {A.~J.~A.}\ \bibnamefont
  {Aquino}}, \bibinfo {author} {\bibfnamefont {H.}~\bibnamefont {Lischka}},
  \bibinfo {author} {\bibfnamefont {F.}~\bibnamefont {Machado}}, \ and\
  \bibinfo {author} {\bibfnamefont {D.}~\bibnamefont {Nachtigallova}},\ }\href
  {\doibase 10.1021/acs.jctc.7b00302} {\bibfield  {journal} {\bibinfo
  {journal} {J. Chem. Theory Comput.}\ }\textbf {\bibinfo {volume} {13}},\
  \bibinfo {pages} {4297} (\bibinfo {year} {2017})}\BibitemShut {NoStop}%
\bibitem [{\citenamefont {Khokhlov}\ and\ \citenamefont
  {Belov}(2020)}]{Khokhlov2020JPCA}%
  \BibitemOpen
  \bibfield  {author} {\bibinfo {author} {\bibfnamefont {D.}~\bibnamefont
  {Khokhlov}}\ and\ \bibinfo {author} {\bibfnamefont {A.}~\bibnamefont
  {Belov}},\ }\href {\doibase 10.1021/acs.jpca.0c01678} {\bibfield  {journal}
  {\bibinfo  {journal} {J. Phys. Chem. A}\ }\textbf {\bibinfo {volume} {124}},\
  \bibinfo {pages} {5790} (\bibinfo {year} {2020})}\BibitemShut {NoStop}%
\end{thebibliography}%

\end{document}